\documentclass[aps, onecolumn, preprint, secnumarabic,nobalancelastpage,nofootinbib]{revtex4-1} 

\usepackage{graphics}      
\usepackage{graphicx}      
\usepackage{longtable}     
\usepackage{url}           
\usepackage{bm,color}            
\usepackage{amssymb, amsmath, enumerate, theorem, epsfig,subfigure}
\usepackage{physics}
\usepackage{verbatim}
\usepackage{upgreek}

\DeclareUnicodeCharacter{0441}{\ensuremath{{}^2}}

\begin{document}

\title{Organic single-photon switch}

\author{Anton Zasedatelev$^{1,*}$, Anton V. Baranikov$^{1}$, Denis Sannikov$^{1}$, Darius Urbonas$^{2}$, Fabio Scafirimuto$^{2}$, Vladislav Yu. Shishkov$^{1,3,4}$, Evgeny S. Andrianov$^{1,3,4}$, Yurii E. Lozovik$^{1,3,4,5}$, Ullrich Scherf$^{6}$, Thilo St\"oferle$^{2}$, Rainer F. Mahrt$^{2}$, and Pavlos G. Lagoudakis$^{1,7,*}$ \footnote{Correspondence address: }}
\date{\today}
\affiliation{$^1$Skolkovo Institute of Science and Technology, Moscow, Russian Federation }
\affiliation{$^2$IBM Research Europe - Zurich, S\"aumerstrasse 4, R\"uschlikon 8803, Switzerland }
\affiliation{$^3$Dukhov Research Institute of Automatics (VNIIA), 22 Sushchevskaya, Mosсow 127055, Russia}
\affiliation{$^4$Moscow Institute of Physics and Technology, 9 Institutskiy per., Dolgoprudny 141700, Moscow region, Russia}
\affiliation{$^5$Institute for Spectroscopy RAS, 5 Fizicheskaya, Troitsk 142190, Russia}
\affiliation{$^6$Macromolecular Chemistry Group and Institute for Polymer Technology, Bergische Universit\"at Wuppertal, Gauss-Strasse 20, 42119 Wuppertal, Germany}
\affiliation{$^7$Department of Physics and  Astronomy, University of Southampton, Southampton, SO17 1BJ, United Kingdom}
{\begin{abstract}{

The recent progress in nanotechnology \cite{Chikkaraddy-Nature-2016, Hail-NatComm-2019} and single-molecule spectroscopy \cite{Maser-NatPhot-2016, Wang-PRX-2017, Wang-NatPhys-2019} paves the way for cost-effective organic quantum optical technologies emergent with a promise to real-life devices operating at ambient conditions. In this letter, we harness $\pi$-conjugated segments of an organic ladder-type polymer strongly coupled to a microcavity forming correlated collective dressed states of light, so-called of exciton-polariton condensates. We explore an efficient way for all-optical ultra-fast control over the macroscopic condensate wavefunction via a single photon. Obeying Bose statistics, exciton-polaritons exhibit an extreme nonlinearity undergoing bosonic stimulation \cite{Zasedatelev-NatPhot-2019} which we have managed to trigger at the single-photon level. Relying on the nature of organic matter to sustain stable excitons dressed with high energy molecular vibrations we have developed a principle that allows for single-photon nonlinearity operation at ambient conditions opening the door for practical implementations like sub-picosecond switching, amplification and all-optical logic at the fundamental limit of single light quanta.

}
								
\end{abstract}  }

\maketitle

Nonlinear optical phenomena are at heart of diverse applications today, facilitating telecommunication, data storage, high resolution microscopy, lithography etc. The implementation of nonlinear effects at single photon level is the holy grail in photonics that drives power consumption of all-optical switches, modulators and transistors towards their fundamental limit, and enables a number of unique applications of optical quantum control for new paradigms of computing, communication and metrology. Accordingly, considerable efforts have been invested toward the achievement of single-photon nonlinearity across a plethora of material systems and physical principles. Platforms range from ultracold atomic ensembles \cite{Peyronel-Nature-2012, Chen-Science-2013, Gorniaczyk-PRL-2014} to a single atom \cite{Reiserer-Science-2013, Shomroni-Science-2014, Tiecke-Nature-2014, Hacker-Nature-2016}, including artificial, solid-state atoms \cite{Volz-NatPhot-2012, Giesz-NatComm-2016, Sun-Science-2018, Dietrich-LasPhotRev-2016} coupled to a high-finesse cavity have been developed addressing this problem. Until recently the nonlinear regime of cavity quantum electrodynamics (cQED) has been uniquely attributed to ultra-cold quantum emitters \cite{Chang-NatPhot-2014}. Recently, organics emerged as a new unique resource for quantum optics, making fascinating cQED phenomena easier to implement at much more relaxed experimental conditions \cite{Wang-PRX-2017, Wang-NatPhys-2019, Ojambati-NatComm-2019}. In this respect, the regime of strong light-matter coupling of a single organic quantum emitter is of particular interest for ambient cQED. Recent progress in this field results in  experimental realization of single molecule vacuum Rabi splitting \cite{Chikkaraddy-Nature-2016, Liu-PRL-2017}, Rabi oscillation and non-classical photon states generated in a strongly-coupled dye molecule at ambient conditions \cite{Ojambati-NatComm-2019}.

Nevertheless, single quantum emitters remain a fragile object to deal with, mainly because they require fine optical and mechanical controls as well as sophisticated  photon detection techniques. Alternatively, the single-photon nonlinearity can be achieved by means of quantum interference in strongly correlated many-body systems, typically based on ultra-cold atomic ensembles  \cite{Peyronel-Nature-2012, Chen-Science-2013, Gorniaczyk-PRL-2014}. Unlike their single emitter counterparts, such systems are more robust and scalable, but sophisticated cooling and states preparation techniques alongside high vacuum conditions compromise their use significantly. In contrast, organic molecular ensembles strongly coupled to a cavity combine the above concepts of cQED and correlated many-body physics giving rise for dressed light-matter states which surpass low temperature limitations \cite{Sanvitto-NatMat-2016}. The dressed states are also known as exciton-polaritons (from here on polaritons), in general. Obeying Bose statistics, under certain conditions polaritons undergo the transition to a macroscopically occupied state exhibiting off-diagonal long-range order - polariton condensate \cite{Deng-RevModPhys-2010, Kasprzak-Nature-2006, Plumhof-NatureMat-2014}. The correlated and highly nonlinear nature of polariton condensates \cite{Carusotto-RevModPhys-2013} makes them outstanding candidates for integrated on-chip photonics \cite{Lerario-NatPhys-2017, Zasedatelev-NatPhot-2019, Sun-NatPhot-2019}. Here, we demonstrate ultra-fast optical switching operational at the fundamental level of single quanta in ambient conditions, using a $\pi$-conjugated ladder-type polymer.

To investigate the potential of polariton condensates for ultra-low energy switching down to a single photon, we employ an organic semiconductor polymer microcavity consisting of a 35 nm thick film of $\pi$-conjugated polymer (MeLPPP) incapsulated in between 50 nm $SiO_{2}$ spacers which is sandwiched between $SiO_{2}/Ta_{2}O_{5}$ distributed Bragg reflectors (DBR) constituting a $\lambda/2$ optical cavity (see Methods). The structure design and related properties have been extensively studied in \cite{Plumhof-NatureMat-2014, Zasedatelev-NatPhot-2019}. Strong light-matter interaction gives rise to new composite states in the system: Frenkel polaritons which obey Bose-Einstein statistics at low density regime. To populate polariton states we first form an exciton reservoir by resonant optical pumping, as shown schematically in Fig.1a, that we direct on the sample at 45 deg incidence to minimise reflection losses. Excitons from the reservoir undergo multiple relaxation processes towards various final states including polaritons through the intracavity radiative pumping and vibron-assisted thermalizaition \cite{Tartakovskii-PRB-2001, Coles-AdvFuncMat-2011, Grant-AOM-2016}. Recently, we have demonstrated the latter favours efficient exciton-to-polariton relaxation in organic systems exhibiting intense vibrational resonances and makes polariton condensation feasible at much lower excitation energy \cite{Zasedatelev-NatPhot-2019}. To exploit high-energy vibrons for exciton cooling we optimise the energy difference between exciton reservoir ($\hslash\omega_{exc}$) and the ground polariton state ($\hslash\omega_{pol}$), thus fulfil the following relaxation condition: $\hslash\omega_{exc}=\hslash\omega_{pol}+\hslash\omega_{vib}$, where $\hslash\omega_{vib}=199meV$ is the molecular vibron energy. By such optimized conditions we reach polariton condensation at low absorbed fluence $F_{th}\sim8\mu J/cm^{2}$ (corresponds to 300 pJ incident pump pulse energy). Above the threshold we observe superlinear increase of total output microcavity emission following stimulated cooling of excitons from the reservoir towards the ground polariton state as shown in Fig.1b.

\begin{figure}[htp]
    \centering
    \includegraphics[width=18.3cm]{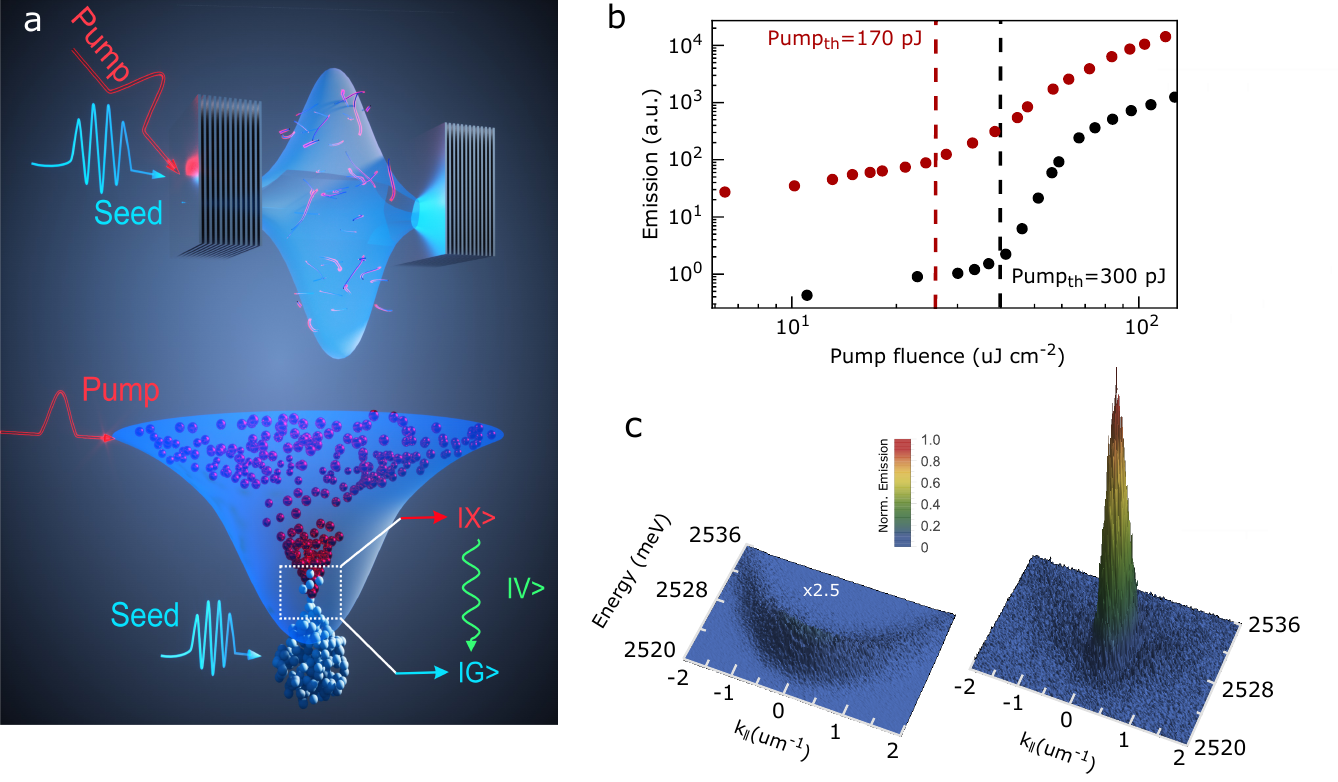}
    \caption{\textbf{The principle of the extreme nonlinearity in organics.} \textbf{a}, Schematic illustrates the principle of stimulated exciton cooling towards the polariton ground state. \textbf{b}, Integrated emission as a function of pump fluence plotted for the spontaneously built (black dots) and seeded (red dots) polariton condensates formed at the pump energy threshold of 300 pJ and 170 pJ respectively; as shown by vertical dashed lines. \textbf{c}, Energy, momentum \textit{(E,k)} distribution for spontaneously built (left) and seeded (right)  polariton condensates respectively. Incident pump fluence is 80 $\mu J/cm^{2}$  ($P\sim 2P_{th}$), energy of the seed beam is 460 aJ.}
    \label{fig:fig1}
\end{figure}

The main idea behind extreme nonlinearity is illustrated in Fig.1a (bottom schematic): few polaritons injected by a seed pulse trigger the exciton avalanche cooling down to the ground polariton state via vibron-assisted bosonic stimulation and result in massively occupied degenerate state forming a polariton condensate. To test the bosonic stimulation principle at ultimately low switching energies of the seed beam we employ a recently developed technique of dynamic polariton condensation \cite{Zasedatelev-NatPhot-2019}. We use spectrally-filtered sub-picosecond long pulses tightly focused on the sample down to 5 $\mu$m spot within $\pm 0.2$ $\mu m^{-1}$  wavevector range seeding the ground polariton state resonantly as schematically depicted in Fig.1a (see Methods). To maximise efficiency of stimulation towards the condensate we optimise spatial and temporal overlap between the seed and pump beams accordingly. Finally, we attenuate the seed down to the level of 460 aJ per pulse corresponding to $\sim1140$ polaritons resonantly injected to the ground state directly. Such a prearrangement in population of the ground polariton state increases the exciton-to-polariton relaxation rate drastically allowing for polariton condensation at the twice lower exciton reservoir density and boosting the total state’s occupancy. According to the pump fluence dependence represented in Fig.1b, resonant seeding results in $\sim50$ times higher integrated emission compared to the spontaneously built condensate. The stimulation effect on energy--momentum \textit{(E,k)} distributions of the condensate is shown in Fig.1c (see Methods).

\begin{figure}[htp]
    \centering
    \includegraphics[width=18.3cm]{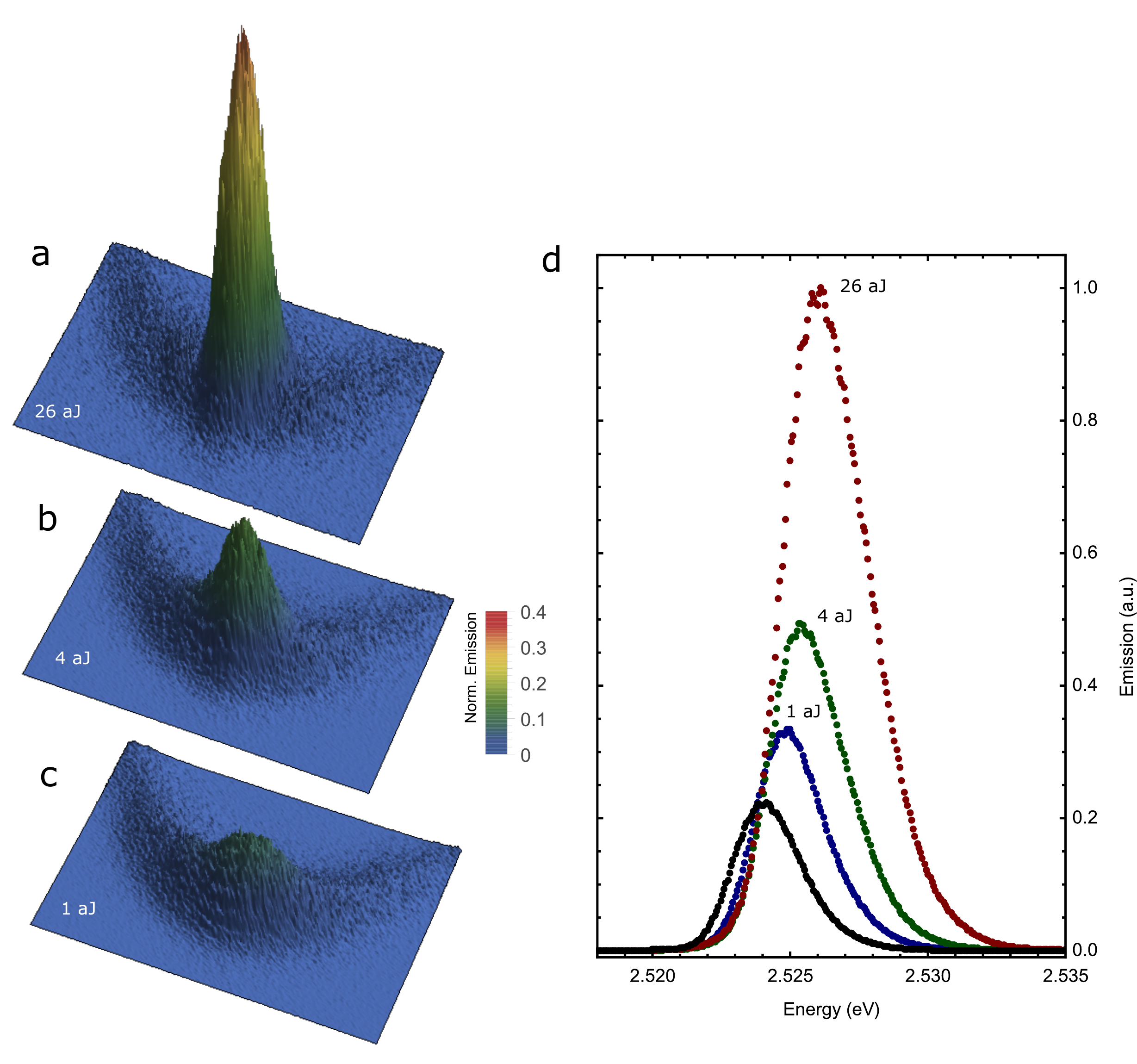}
    \caption{\textbf{Atto-Joule polariton switch.} \textbf{a,b,c}, Energy, momentum \textit{(E,k)} distributions of 5000 condensates realizations each seeded with 26 aJ, 4 aJ and 1 aJ pulses respectively. \textbf{d}, Emission spectra obtained from \textit{(E,k)} distributions \textbf{a-c} by integrating over $\pm 0.2$ $\mu m^{-1}$ in momentum space.  Red, green and blue dots correspond to the polariton condensate seeded with 26 aJ, 4 aJ and 1 aJ pulses respectively, while black dots display emission spectrum of the spontaneously built condensate. }
    \label{fig:fig2}
\end{figure}

Started out from the sub-femtojoule seed pulses we further investigate potential control over the condensate occupancy by harnessing progressively smaller seed energy: from 26 aJ down to 4 aJ and 1 aJ pulses as shown in Fig.2a-c respectively. With decreasing energy of the seed we observe substantial reduction of the ground state population, although even at the lowest seed energy of 1 aJ carrying only 2.5 photons per pulse on average we are still able to resolve an increase in the condensate population. Figure 2d shows emission spectra obtained from integrated \textit{E,k} distributions over the momentum range of $\pm 0.2$ $\mu m^{-1}$, exhibiting a clear contrast in population of seeded and spontaneously formed condensates. Comparing the spectral area in Fig.2d we detect nearly 55\% difference in the ground state population for the case of 1 aJ seed. To minimize fluctuations of the condensate occupation number caused by pump induced exciton reservoir density variations we excite the system close to the gain saturation regime at $2F_{th}$ and integrate the output signal over 5000 single condensate realizations on the detector (see Methods).

To quantify the condensate nonlinearity upon seeding photon states, we introduce the contrast as the figure of merit parameter which relies on the ratio of integrated population at the ground state between spontaneously formed ($P_{spont}$) and seeded polariton condensates ($P_{seed}$): $Contrast = \frac{P_{seed}}{P_{spont}}-1 $. Our systematic study of the seeded condensation reveals nearly a power-law dependence on the average seed photon states as follows from the experimental data represented in Fig.3. We experimentally observe condensate switching triggered by the seed pulse carrying $\sim1.5$ photons on average with a probability larger than 0.95. Figure 3 shows that the lowest limit of $2\sigma$ error bars exceeds 0\% contrast level of unseeded condensate. To explain the experimentally observed behaviour we have developed a microscopic theory of the seeded vibron-mediated polariton condensation (see Methods and Sections I in Supplementary Information (SI) for further details). Numerical simulations agree well with the observed contrast dependence, as one can see from the fit in Fig.3 (solid curve). In-depth analysis of  mechanisms behind the nonlinearity is represented in Section II of SI. It is worth noting, the principle of switching allows for single photon operation as follows from the simulations. Our model predicts a contrast $\sim20\%$ for single-photon seed.

However, despite the good agreement between experimental data and theory care has to be taken in approaching the single photon regime experimentally as the statistical properties of the source can no longer be neglected. The seed, being a laser-like source, obeys Poisson photon number distribution inherited from the laser generating the supercontinuum beam (see Methods). Therefore at the few-photon regime, the tail of the distribution containing higher photon number states might contribute to polariton stimulation significantly. Following Poisson statistics, the probability of having \textit{n} photons in the distribution scales with its maximum at \textit{$\langle n \rangle$} average photons as $\frac{{\langle n \rangle}!}{n!}{\langle n \rangle}^{n-{\langle n \rangle}}$. In the case of a seed pulse with 4 photons on average, although the majority of realisations exhibits 4 photons per pulse, there are some relatively rare events containing significantly more photons per pulse. For instance 10-photon events are possible but with $\sim37$ times lower probability. Since in the above experiments we integrate over thousands of single-condensate realizations our results might be affected by such higher photon number states, providing a stronger stimulation rate and being accumulated with a large weight in the total registered polariton population. In other words it is not clear now whether the few-photon control over the condensate is experimentally feasible by every seed pulse or cumulative effect is observed.

\begin{figure}[htp]
    \centering
    \includegraphics[width=8.9cm]{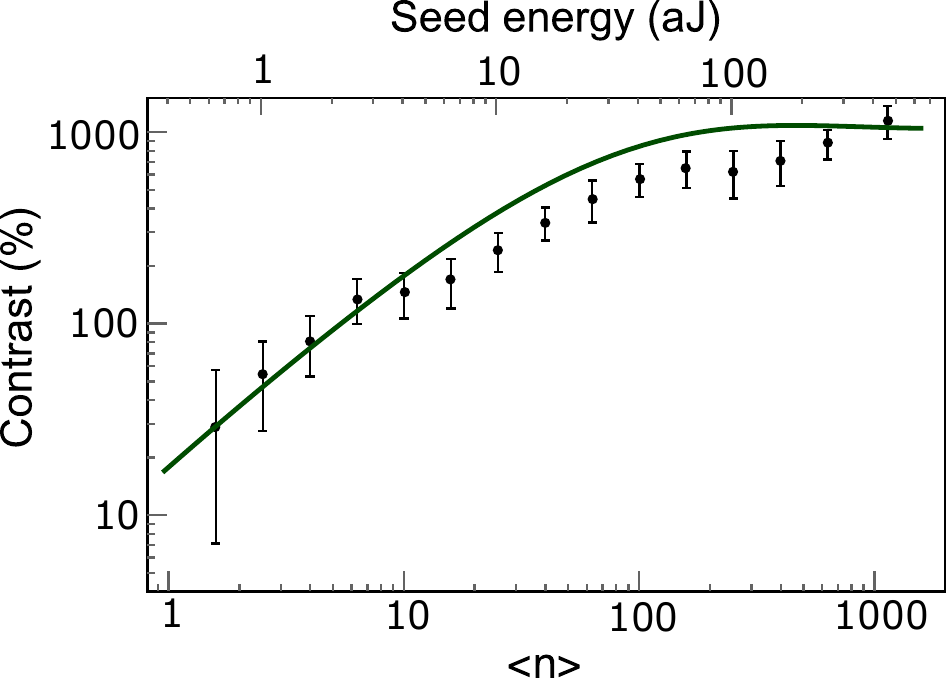}
    \caption{\textbf{The switching contrast at the single-photon level.} The contrast in population between seeded and spontaneously built polariton condensates as the function of seed energy and average photon number per pulse. Solid curve shows the result of numerical simulations carried out according with the theoretical model.}
    \label{fig:fig3}
\end{figure}

We address this problem by implementing a single-shot measurement scheme allowing for pulse-to-pulse analysis of the condensate population (see Methods). We employ the same pumping and seeding excitation geometries but we detect condensate population in real space using appropriate filtering in momentum space within $\pm1 \mu m^{-1}$ (see Methods). We perform these measurements by seeding the ground state for 300 sequential pulses then we switch the seed beam off allowing for 300 realizations of spontaneously built polariton condensates and repeat the whole sequence again as shown in Fig.4a-e. A clear contrast for single-condensate realisations with an average value of about 350\%, 160\%, 60\%, 25\% and 11\% assessed for the seed containing $\langle n \rangle$ equals 600, 60, 9.3, 2.7, and 1 photon per pulse respectively can be seen. Statistical analysis of 900 realizations elucidates the effect of single-photon control over the total polariton population at the condensate induced by every seed pulse. Right panels of Fig.4a-e show histograms of the total polariton population at single spontaneously built (black) and seeded (red) condensate realizations fitted by Gaussian distributions. As the experimental distributions are close to Gaussian our statistical analysis yields almost the same mean contrast values of 360\%, 160\%, 63\%, 26\% and 11\% as shown in Fig.4f. Moreover, the results of single condensate realisations are consistent with integrated measurements discussed above and are well supported by the theory.

\begin{figure}[htp]
    \centering
    \includegraphics[width=18.3cm]{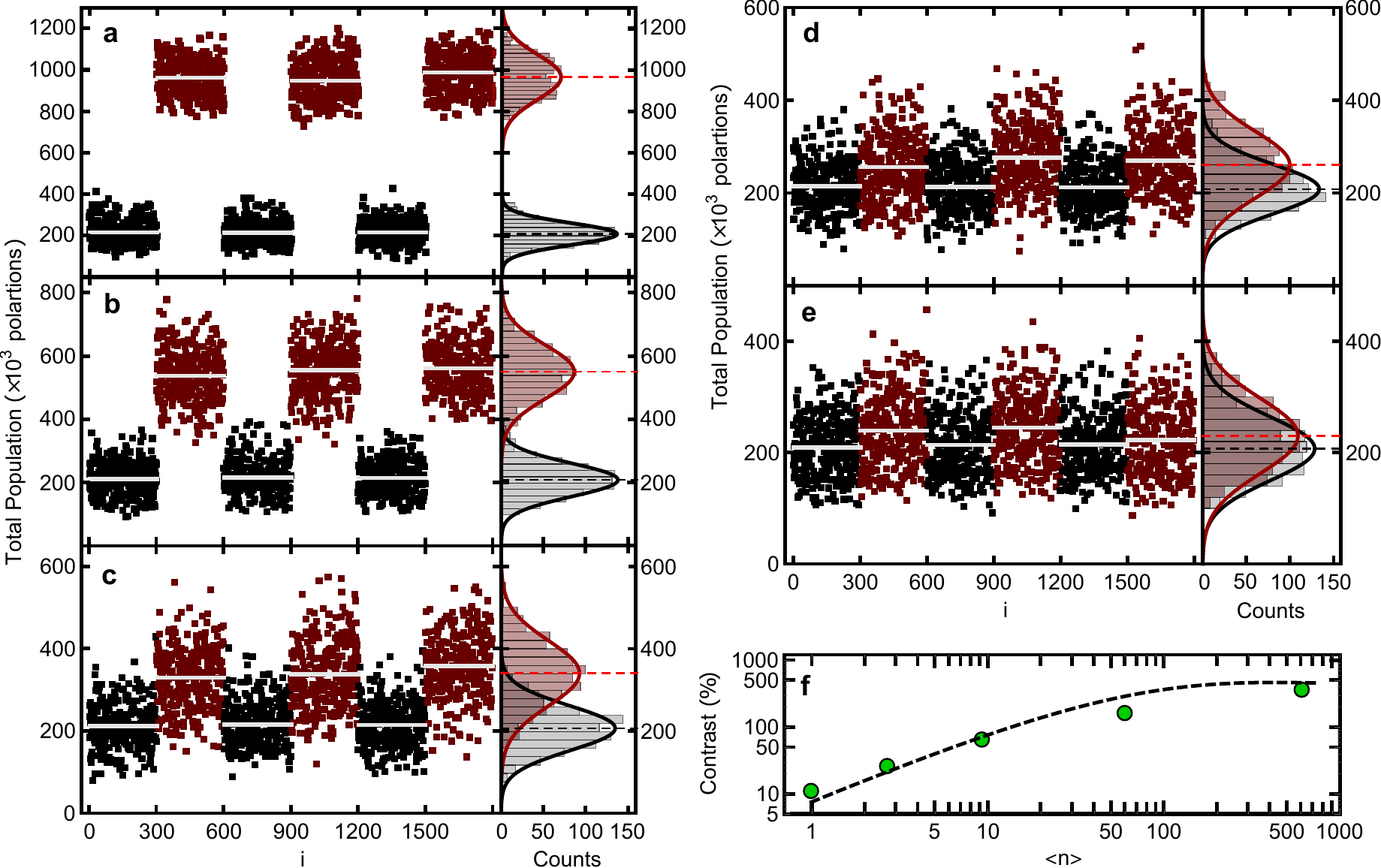}
    \caption{\textbf{Single-photon switching at single-condensate realisations.} \textbf{a-e}, Total polariton populations recorded within $\pm1 \mu m^{-1}$ wavevector range in the single-shot regime for a sequence of 300-to-300 spontaneously built (black dots) and seeded (red dots) polariton condensate realisations. Seeded realisations are triggered by $\langle n \rangle$ equals 600, 60, 9.3, 2.7, and 1 photon per pulse on average as shown in \textbf{a-e} respectively. Right panels represent histograms of total polariton population within single-condensate realisations seeded with correspondent $\langle n \rangle$ photon number state (red) as well histograms of the spontaneously built condensate (black), both fitted by Gaussian distributions accordingly. \textbf{f}, Switching contrast versus average number of photons per seed pulse $\langle n \rangle$ acquired from statistical analysis according to the histograms. }
    \label{fig:fig4}
\end{figure}

In conclusion, we exemplify the outstanding single-photon nonlinearity discovered in organic polariton condensates. The general mechanism that enables nonlinearity at the quantum limit is the bosonic stimulation principle inherited for exciton-polartions regardless of the material systems. Our report convincingly demonstrates few-photon control over the polariton occupancy of the condensate with unprecedented gain of $\sim23000$ secondary polaritons per single resonantly injected seed polariton. We experimentally demonstrate for the first time at ambient conditions single-photon polariton condensate ultrafast switching with a contrast of $\sim11\%$ following the pulse-to-pulse analysis of the single-condensate realisations and $\sim20\%$ according to integrated measurements. Organic materials offer great potential for integration on a chip together with their identical chemical and optical characteristics which can be easily altered spanning a range from ultraviolet to the near infrared spectrum. Ultimately, it opens a new horizon for all-optical manipulation, data processing and detection at the single-photon level bridging quantum properties of light with classical phenomena of massively occupied states such as mixed light-matter Bose-Einstein condensates.

\clearpage

\section*{Methods}

\textbf{Sample fabrication}.

The sample is composed of a bottom distributed Bragg reflector (DBR), a central cavity defect region with an effective thickness slightly larger than half the exciton wavelength, and a top DBR on a fused silica substrate. The DBRs consist of alternating SiO$_{2}$/Ta$_{2}$O$_{5}$ quarter-wavelength-thick layers produced by sputter deposition (9+0.5 pairs for the bottom DBR, 6+0.5 for the top DBR). The center of the cavity consists of a polymer layer sandwiched within 50-nm spacer layers of sputtered SiO$_{2}$. The SiO$_{2}$ spacer is sputtered on the organic using a SiO$_{2}$ sputter target. Methyl-substituted ladder-type poly(p-phenylene) (MeLPPP; $M_{n}=31500$, $M_{w}=79000$) was synthesized as described elsewhere \cite{Scherf-Macromolecularchemistry-1992}. MeLPPP is dissolved in toluene and spin-coated on the bottom spacer layer. The film thickness of approximately 35 nm is measured with a profilometer (Veeco Dektak).

\textbf{Spectroscopy}.
The pump beam with 150-200 fs pulse duration was generated by a tunable optical parametric amplifier (Coherent OPerA SOLO) which was pumped by 500 Hz high energy Ti:Sapphire regenerative amplifier (Coherent Libra-HE). The center wavelength was adjusted with respect to experiments at 2.72 eV having 30 meV full-width at half-maximum (FWHM). The pulses were focused at the sample to a 20 $\mathrm{\upmu{}m}$ spot and hit the sample at 45$^{\circ}$. Filtered broadband white-light-continuum (WLC) generated in a sapphire plate with photon energies in the range 2.45 - 2.6 eV was utilized as the seed beam having 150-250 fs pulse duration. The seed beam was focused on the sample at the normal incidence to 5 $\mathrm{\upmu{}m}$ spot size with a micro-objective (10x Nikon, 0.3 NA) which allows to seed the ground polariton state within a wavevector range of $\pm 0.2 \mu m^{-1}$. In all measurements, temporal and spatial overlapping between the control and the pump beams were optimized by maximizing the signal of the output nonlinear emission from the sample. Temporal overlapping was varied by using a motorized translation stage with retroreflector in the seed beam optical path.

Integrated energy-momentum \textit{E,k} distributions were acquired in transmission geometry. Output emission of the sample was collected with a 10X Mitutoyo plan apo infinity corrected objective (NA = 0.28) and coupled to a 750 mm spectrometer (Princeton Instruments SP2750) equipped with an electron multiplying CCD camera (Princeton Instruments ProEM 1024BX3). The emission was spectrally and in-plane wavevector resolved using a 1200 grooves/mm grating and a slit width of 50 $\mathrm{\upmu{}m}$ at the entrance of the spectrometer. To obtain the incident excitation density of the pump pulse, the average pump power was measured using a calibrated Si photodetector (Thorlabs-Det10/M) and an oscilloscope (Keysight DSOX3054T) for data acquisition. Accuracy verification of the power measurements was carried out by using two independent power meters: 1 - Si photodiode power sensor (Thorlabs-S120VC) with a power meter console (Thorlabs-PM100D), 2 - thermal power sensor (Thorlabs-S302C) equipped with the power meter console (Thorlabs-PM100D). The energy of the seed beam was calibrated by using standard power meter measurements and additionally verified through a single photon counting technique using a photon counting module (SPC-160, Becker \& Hickl GMBH) and a single-photon avalanche Si photodiode (IDQ 100), for details see Fig.S4 in Section III of SI. All the measurements for Fig.2 and 3 are carried out under the same incident pump fluence of $F_{pump}=80 \mu J cm^{-2}$ ($P\sim 2P_{th}$) to minimize noise level originating from pump power fluctuations and high nonlinearity of the sample above the threshold. Pump fluence dependencies in Fig.1b as well as spectra in Fig.2d and the contrast dependence in Fig.3 are plotted for the ground polariton state integrated within $\pm 0.2 \mu m^{-1}$. 

Polariton single-condensate realizations were investigated by means of single-shot real space imaging technique using the same pump conditions ($P\sim 2P_{th}$) with individual pulse control. The images were recorded by means of the electron multiplying CCD camera applying 100x EM gain and operating in a single frame acquisition mode. To get rid of non-condensed polariton density we filtered out the output emission above $\pm 1 \mu m^{-1}$ in Fourier space. The single-condensate realisations in Fig.4a-e were recorded by seeding the ground state for 300 sequential pulses followed by seed beam off measurements that allow for 300 realizations of spontaneously built polariton condensates. The whole sequence was repeated three times for each seed pulse energy. The incident seed energy was stabilised at the level of 2\% standard deviation. To reduce the noise originated from pump fluctuation we recorded the energy of incident pulses for each condensate realisation using calibrated Si photodetector and an oscilloscope and then extracted only realisations within 2\% of tolerance interval. Statistical analysis and calculation of the single-realisation contrast in Fig.4a-e were carried out through processing of all 900 spontaneously built and 900 seeded condensate realisations for each seed state.

\clearpage

\textbf{Theoretical model}.

The system is treated microscopically by considering a thin organic active layer (MeLPPP) placed in the cavity excited by an external wave that corresponds to the pump beam. The pump wave induces the dipole moment transition of organics creating Frenkel-type excitons. In turn, the excitons interacts with both vibrons and cavity modes. Hereby we consider the pump exciting organic layer resonantly to the exciton transition. The Hamiltonian of the whole system has the following form:
\begin{equation} \label{FullHamiltonianM}
\hat H = {\hat H_{{\rm{cav}}}} + {\hat H_{{\rm{exc}}}} + {\hat H_{{\rm{vib}}}} + {\hat H_{{\rm{exc - cav}}}} + {\hat H_{{\rm{exc - vib}}}}
 + {\hat H_{{\rm{exc - pump}}}} + {\hat H_{{\rm{cav - seed}}}}
\end{equation}
where the terms correspond to uncoupled cavity modes, excitons, vibrons, interactions between excitons and cavity modes, excitons and vibrons as well as interaction between excitons and external pump wave and cavity modes with the external seed wave, listed as they appear in Eq.~(\ref{FullHamiltonianM}). For further details see Section I in SI.
On the next step we make a transformation from treating the system in form of uncoupled excitons and cavity modes to their hybridized solutions: exciton-polaritons. We diagonalize the corresponding part of the full Hamiltonian, namely ${\hat H_{{\rm{cav}}}} + {\hat H_{{\rm{exc}}}} + {\hat H_{{\rm{exc - cav}}}}$, and derive operators for lower polaritons ${\hat s_{{\rm{low}}{\bf{k}}}}$ and upper polaritons ${\hat s_{{\rm{up}}{\bf{k}}}}$ using the transformation below:
\begin{equation} \label{TransformationForUpperPolaritonsM}
{\hat s_{{\rm{up}}{\bf{k}}}} = {\hat c_{{\rm{exc}}{\bf{k}}}}\cos {\varphi _{\bf{k}}} + {\hat a_{{\rm{cav}}{\bf{k}}}}\sin {\varphi _{\bf{k}}}
\end{equation}
\begin{equation} \label{TransformationForLowerPolaritonsM}
{\hat s_{{\rm{low}}{\bf{k}}}} = {\hat a_{{\rm{cav}}{\bf{k}}}}\cos {\varphi _{\bf{k}}} - {\hat c_{{\rm{exc}}{\bf{k}}}}\sin {\varphi _{\bf{k}}}
\end{equation}
where ${\hat a_{{\rm{cav}}{\bf{k}}}}$ and ${\hat c_{{\rm{exc}}{\bf{k}}}}$ are cavity photon and exciton operators respectively, and
\begin{equation} \label{TransformationAngleM}
{\varphi _{\bf{k}}}=
\frac{1}{2}{\rm arctg}\left(       
\frac{2 \Omega_{{\rm R}{\bf k}}}{\omega _{{\rm{exc}}} - \omega _{{\rm{cav}}{\bf{k}}}}
 \right)
\end{equation}
 Following the transformations above the full Hamiltonian~(\ref{FullHamiltonianM}) takes the form
\begin{multline} \label{TransHamiltonianM}
\hat H = \sum\limits_{\bf{k}} {\hbar {\omega _{{\rm{up}}{\bf{k}}}}\hat s_{{\rm{up}}{\bf{k}}}^\dag {{\hat s}_{{\rm{up}}{\bf{k}}}}}  + \sum\limits_{\bf{k}} {\hbar {\omega _{{\rm{low}}{\bf{k}}}}\hat s_{{\rm{low}}{\bf{k}}}^\dag {{\hat s}_{{\rm{low}}{\bf{k}}}}}  + \sum\limits_{\bf{q}} {\hbar {\omega _{{\rm{vib}}}}\hat b_{\bf{q}}^\dag {{\hat b}_{\bf{q}}}}  + \\
 + \sum\limits_{\bf{k}} {\sum\limits_{{\bf{k'}}} {\hbar {g}\left( {\hat s_{{\rm{up}}{\bf{k}}}^\dag \cos {\varphi _{\bf{k}}} - \hat s_{{\rm{low}}{\bf{k}}}^\dag \sin {\varphi _{\bf{k}}}} \right)\left( {{{\hat s}_{{\rm{up}}{\bf{k'}}}}\cos {\varphi _{{\bf{k'}}}} - {{\hat s}_{{\rm{low}}{\bf{k'}}}}\sin {\varphi _{{\bf{k'}}}}} \right)\left( {\hat b_{ - \left( {{\bf{k}} - {\bf{k'}}} \right)}^\dag  + {{\hat b}_{{\bf{k}} - {\bf{k'}}}}} \right)} }  + \\
 + \sum\limits_{\bf{k}} {\hbar {\Omega _{\bf{k}}}\left( t \right)\left( {\left( {\hat s_{{\rm{up}}{\bf{k}}}^\dag \cos {\varphi _{\bf{k}}} - \hat s_{{\rm{low}}{\bf{k}}}^\dag \sin {\varphi _{\bf{k}}}} \right){e^{ - i{\omega _\Omega }t}} + \left( {{{\hat s}_{{\rm{up}}{\bf{k'}}}}\cos {\varphi _{{\bf{k'}}}} - {{\hat s}_{{\rm{low}}{\bf{k'}}}}\sin {\varphi _{{\bf{k'}}}}} \right){e^{i{\omega _\Omega }t}}} \right)}  + \\
 + \sum\limits_{\bf{k}} {\hbar {W_{\bf{k}}}\left( t \right)\left( {\left( {\hat s_{{\rm{up}}{\bf{k}}}^\dag \cos {\varphi _{\bf{k}}} - \hat s_{{\rm{low}}{\bf{k}}}^\dag \sin {\varphi _{\bf{k}}}} \right){e^{ - i{\omega _W}t}} + \left( {{{\hat s}_{{\rm{up}}{\bf{k'}}}}\cos {\varphi _{{\bf{k'}}}} - {{\hat s}_{{\rm{low}}{\bf{k'}}}}\sin {\varphi _{{\bf{k'}}}}} \right){e^{i{\omega _W}t}}} \right)}
\end{multline}
 The polaritons and vibrons interact with the environment that in turn initiates relaxation processes. We describe the relaxation processes within the formalism of Lindblad superoperators for a general density matrix $\hat \rho$ of both polaritons and vibrons. The Lindblad superoperators for the lower, upper polaritons and vibrons respectively have the form
\begin{equation} \label{LindbladDissipatonLowerPolaritonsM}
{L_{{\rm{low}}}}\left( {\hat \rho } \right) = \sum\limits_{\bf{k}} {\frac{{{\gamma _{{\rm{low}}{\bf{k}}}}}}{2}\left( {2{{\hat s}_{{\rm{low}}{\bf{k}}}}\hat \rho \hat s_{{\rm{low}}{\bf{k}}}^\dag  - \hat s_{{\rm{low}}{\bf{k}}}^\dag {{\hat s}_{{\rm{low}}{\bf{k}}}}\hat \rho  - \hat \rho \hat s_{{\rm{low}}{\bf{k}}}^\dag {{\hat s}_{{\rm{low}}{\bf{k}}}}} \right)}
\end{equation}

\begin{equation} \label{LindbladDissipatonUpperPolaritonsM}
{L_{{\rm{up}}}}\left( {\hat \rho } \right) = \sum\limits_{\bf{k}} {\frac{{{\gamma _{{\rm{up}}{\bf{k}}}}}}{2}\left( {2{{\hat s}_{{\rm{up}}{\bf{k}}}}\hat \rho \hat s_{{\rm{up}}{\bf{k}}}^\dag  - \hat s_{{\rm{up}}{\bf{k}}}^\dag {{\hat s}_{{\rm{up}}{\bf{k}}}}\hat \rho  - \hat \rho \hat s_{{\rm{up}}{\bf{k}}}^\dag {{\hat s}_{{\rm{up}}{\bf{k}}}}} \right)}
\end{equation}

\begin{equation} \label{LindbladDissipatonVibronsM}
{L_{{\rm{vib}}}}\left( {\hat \rho } \right) = \sum\limits_{\bf{q}} {\frac{{{\gamma _{{\rm{vib}}}}}}{2}\left( {1 + n_{\rm{vib}}^{{\rm{th}}}} \right)\left( {2{{\hat b}_{\bf{q}}}\hat \rho \hat b_{\bf{q}}^\dag  - \hat b_{\bf{q}}^\dag {{\hat b}_{\bf{q}}}\hat \rho  - \hat \rho \hat b_{\bf{q}}^\dag {{\hat b}_{\bf{q}}}} \right)}
 + \sum\limits_{\bf{q}} {\frac{{{\gamma _{{\rm{vib}}}}}}{2}n_{\rm{vib}}^{{\rm{th}}}\left( {2\hat b_{\bf{q}}^\dag \hat \rho {{\hat b}_{\bf{q}}} - {{\hat b}_{\bf{q}}}\hat b_{\bf{q}}^\dag \hat \rho  - \hat \rho {{\hat b}_{\bf{q}}}\hat b_{\bf{q}}^\dag } \right)}
\end{equation}
where ${\gamma _{{\rm{low}}{\bf{k}}}}$, ${\gamma _{{\rm{up}}{\bf{k}}}}$ and ${\gamma _{{\rm{vib}}}}$ are the dissipation rates of lower, upper polaritons and the vibrons respectively, and $n_{\rm{vib}}^{{\rm{th}}} = {1 \mathord{\left/
 {\vphantom {1 {\left( {\exp \left( {{{\hbar {\omega _{{\rm{vib}}}}} \mathord{\left/
 {\vphantom {{\hbar {\omega _{{\rm{vib}}}}} T}} \right.
 \kern-\nulldelimiterspace} T}} \right) - 1} \right)}}} \right.
 \kern-\nulldelimiterspace} {\left( {\exp \left( {{{\hbar {\omega _{{\rm{vib}}}}} \mathord{\left/
 {\vphantom {{\hbar {\omega _{{\rm{vib}}}}} T}} \right.
 \kern-\nulldelimiterspace} T}} \right) - 1} \right)}}$ is the thermal distribution of vibrons at an effective MeLPPP temperature $T$.
 
Thermalization of the lower polaritons is described by the following Lindblad term:

\begin{equation} \label{LindbladThermalizationLowerPolaritonsM}
{\hat L_{{\rm{th}}}}\left( {\hat \rho } \right) = \sum\limits_{{{\bf{k}}_1},{{\bf{k}}_2}} {\frac{{\gamma _{{\rm{low}}}^{{{\bf{k}}_1}{{\bf{k}}_2}}}}{2}\left( {2{{\hat s}_{{\rm{low}}{{\bf{k}}_2}}}\hat s_{{\rm{low}}{{\bf{k}}_1}}^\dag \hat \rho {{\hat s}_{{\rm{low}}{{\bf{k}}_1}}}\hat s_{{\rm{low}}{{\bf{k}}_2}}^\dag  - {{\hat s}_{{\rm{low}}{{\bf{k}}_1}}}\hat s_{{\rm{low}}{{\bf{k}}_2}}^\dag {{\hat s}_{{\rm{low}}{{\bf{k}}_2}}}\hat s_{{\rm{low}}{{\bf{k}}_1}}^\dag \hat \rho  - \hat \rho {{\hat s}_{{\rm{low}}{{\bf{k}}_1}}}\hat s_{{\rm{low}}{{\bf{k}}_2}}^\dag {{\hat s}_{{\rm{low}}{{\bf{k}}_2}}}\hat s_{{\rm{low}}{{\bf{k}}_1}}^\dag } \right)} 
\end{equation}
Here $\gamma _{{\rm{low}}}^{{{\bf{k}}_1}{{\bf{k}}_2}}$ is the rate of energy flow from the lower polaritons having in-plane wavevector ${{\bf{k}}_2}$ towards ones with  wavevector ${{\bf{k}}_1}$. According to  Kubo-Martin-Schwinger relation, the rate $\gamma _{{\rm{low}}}^{{{\bf{k}}_1}{{\bf{k}}_2}}$ satisfies 

\begin{equation} \label{KMShLowerThermalizationRatesM}
\gamma _{{\rm{low}}}^{{{\bf{k}}_1}{{\bf{k}}_2}} = \gamma _{{\rm{low}}}^{{{\bf{k}}_2}{{\bf{k}}_1}}\exp \left( {\frac{{{\omega _{{\rm{low}}{{\bf{k}}_2}}} - {\omega _{{\rm{low}}{{\bf{k}}_1}}}}}{T}} \right)
\end{equation}

where $T$ is the effective lattice temperature.

Finally, we use a standard Lindblad master equations including the full Hamiltonian~(\ref{TransHamiltonianM}) and the Lindblad terms~(\ref{LindbladDissipatonLowerPolaritonsM}-\ref{LindbladThermalizationLowerPolaritonsM}) described above: 

\begin{equation} \label{MasterEquationM}
{\frac{ d\hat \rho}{dt}} = {\frac{ i}{\hbar}}{\left[ {\hat \rho}, {\hat H} \right]}
+ {\hat L_{{\rm{up}}}}\left( {\hat \rho } \right)
+ {\hat L_{{\rm{low}}}}\left( {\hat \rho } \right)
+ {\hat L_{{\rm{vib}}}}\left( {\hat \rho } \right)
+ {\hat L_{{\rm{th}}}}\left( {\hat \rho } \right)
\end{equation}

For further details see Sections I and II in SI.

\clearpage

\section*{Data availability}
All data supporting this study are openly available from the University of Southampton repository at https://doi.org/10.5258/SOTON/D1374.

\section*{Acknowledgements}
This work was supported by the Russian Scientific Foundation (RSF) grant No. 20-72-10145 and the UKs Engineering and Physical Sciences Research Council grant EP/M025330/1 to P.G.L on Hybrid Polaritonics. E.S.A. and V.Yu.Sh. thank the Foundation for the Advancement of Theoretical Physics and Mathematics Basis. Yu.E.L. acknowledges support by Program of Basic Research of the National Research University Higher School of Economics. D.U., F.S. and T.S. acknowledge support by QuantERA project RouTe (SNSF Grant No. 20QT21 175389). P.G.L, D.U., T.S. and R.F.M. acknowledge support by European H2020-FETOPEN project POLLOC (Grant No. 899141).

\section*{Author contributions}
A.Z., A.V.B. and D.S. performed the experiments and analysed the data. D.U., F.S., T.S., and R.F.M. contributed to the design and fabrication of the organic microcavity. U.S. synthesised the organic material. V.Yu.Sh., E.S.A. and Yu.E.L. developed microscopic theory and carried out numerical simulations. A.Z. and P.G.L. designed and led the research. The manuscript was written through contributions from all authors. All authors have given approval to the final version of the manuscript.

\section*{Additional information}
The authors declare no competing financial interests.

\pagebreak

\renewcommand{\arraystretch}{1.5} 
 \setcounter{figure}{0}
\renewcommand{\thefigure}{S\arabic{figure}}
 \setcounter{table}{0}
\renewcommand{\thetable}{S\arabic{table}}

{\bf{\large{Supplementary Information: Organic single-photon switch}}}

\setcounter{equation}{0}

\section{Microscopic theory of vibron-mediated polariton condensation}

We consider a nonequilibrium microscopic model which describes an ensemble of organic molecules with vibrationally dressed electronic transitions and coupled to cavity modes carrying different in-plane momenta $\hbar{}\bf{k}_{\parallel}$ (hereinafter $\hbar{}{\bf{k}}$). The electronic transition being localised at molecules are effectively treated as Frenkel-type excitons, where each excitation can be described by the Pauli creation and annihilation operators acting on a single molecule. As the system excited in resonance with excitons (see Methods) we introduce an extra term coupling excitons with an external pump electromagnetic field. In addition, we implement the coupling term for a weak electromagnetic field resonantly seeding the ground polariton state of the system. Hereby, the full Hamiltonian of the systems reads:
\begin{equation} \label{FullHamiltonian}
\hat H = {\hat H_{{\rm{cav}}}} + {\hat H_{{\rm{exc}}}} + {\hat H_{{\rm{vib}}}} + {\hat H_{{\rm{exc - cav}}}} + {\hat H_{{\rm{exc - vib}}}}
 + {\hat H_{{\rm{exc - pump}}}} + {\hat H_{{\rm{cav - seed}}}}
\end{equation}

The Hamiltonian of the cavity ${\hat H_{{\rm{cav}}}}$ is
\begin{equation} \label{CavityHamiltonian}
{\hat H_{{\rm{cav}}}} = \sum\limits_{\bf{k}} {\hbar {\omega _{{\rm{cav}}{\bf{k}}}}\hat a_{{\rm{cav}}{\bf{k}}}^\dag {{\hat a}_{{\rm{cav}}{\bf{k}}}}}
\end{equation}
where $\hat a_{{\rm{cav}}{\bf{k}}}^\dag$ and ${\hat a_{{\rm{cav}}{\bf{k}}}}$ are the creation and annihilation operators for a photon in the cavity which obey commutation relation $\left[ {{{\hat a}_{{\rm{cav}}{\bf{k}}}},\hat a_{{\rm{cav}}{\bf{k'}}}^\dag } \right] = {\delta _{{\bf{k}},{\bf{k'}}}}$, the wavevector ${\bf{k}}$ corresponds to in-plane momentum $\hbar{}{\bf{k}}$, and ${\omega _{{\rm{cav}}{\bf{k}}}}$ is the eigenfrequency of the cavity mode with the in-plane wavevector ${\bf{k}}$.

The Hamiltonian of the excitons ${\hat H_{{\rm{exc}}}}$ is
\begin{equation} \label{ExcitonsHamiltonianRealSpace}
{\hat H_{{\rm{exc}}}} = \sum\limits_{j} {\hbar {\omega _{{\rm{exc}}}}\hat \sigma_{{\rm{exc}}{j}}^\dag {{\hat \sigma}_{{\rm{exc}}{j}}}}
\end{equation}
where ${\omega _{{\rm{exc}}}}$ is the eigenfrequency of the excitons, $\hat \sigma_{{\rm{exc}}{j}}^\dag$ and ${\hat \sigma_{{\rm{exc}}{j}}}$ are the creation and annihilation operators of the exciton of a single $\pi$--conjugated segment of MeLPPP located at the point ${{\bf r}_j}$ . These operators obey commutation relation ${{\hat \sigma}_{{\rm{exc}}{j}}} {{\hat \sigma}_{{\rm{exc}}{j'}}^\dag} + {{\hat \sigma}_{{\rm{exc}}{j'}}^\dag} {{\hat \sigma}_{{\rm{exc}}{j}}}  = {\delta _{{j},{j'}}}$. Below we consider the case of small probability for the exciton to be found in an excited state. In this case the approximate commutation relation $\left[ {{{\hat \sigma}_{{\rm{exc}}{j}}},\hat \sigma_{{\rm{exc}}{j'}}^\dag } \right] \approx {\delta _{{j},{j'}}}$ is valid. 

The Hamiltonian of interaction between excitons and cavity modes ${\hat H_{{\rm{exc - cav}}}}$ is a multi-mode Jaynes-Cummings Hamiltonian \cite{ScullyZubairy1997}
\begin{equation} \label{ExcitonsCavityHamiltonianRealSpace}
{\hat H_{{\rm{exc - cav}}}} = 
\sum\limits_{{\bf k}, j} {\hbar {\Omega _{{\rm{1R}}{{\bf k}}}}\left( 
{\hat \sigma_{{\rm{exc}}{j}}^\dag {{\hat a}_{{\rm{cav}}{\bf{k}}}}{e^{i{\bf k}{\bf r}_j}} 
+ {{\hat \sigma}_{{\rm{exc}}{j}}}\hat a_{{\rm{cav}}{\bf{k}}}^\dag {e^{-i{\bf k}{\bf r}_j}} } 
\right)}
\end{equation}
where ${\Omega _{{\rm{1R}}{\bf{k}}}}$ is the Rabi frequency of interaction between an exciton placed at ${{\bf r}_j}$ and a cavity mode with the in-plane wavevector ${\bf{k}}$. Here we suppose that electric field of the $\bf{k}$th mode is distributed in plane parallel to the mirrors according to $e^{i \bf{kr}}$ and constant in the anti-node position where the MeLPPP layer is located. It is reasonable approximation since we deal with the microcavity carrying a single fundamental $\lambda/2$ mode and containing 35 nm thin MeLPPP layer at the center, which is much smaller than the wavelength $\lambda$. The explicit expression for the Rabi frequency ${\Omega _{{\rm{1R}}{\bf{k}}}}$ is ${\Omega _{{\rm{1R}}{\bf{k}}}} =  - {{\bf{E}}_{\bf{k}}}{\bf{d}}/\hbar $, where $\bf{d}$ is the transition dipole momentum of the exciton and ${ E}_{\bf k}$ is the electric field amplitude per "one photon" in the cavity with the in-plane momentum $\hbar{}{\bf{k}}$.

It is convenient to introduce the collective operator effectively describing all excitons, namely,
\begin{equation} \label{ExcitonsFourier}
 {\hat c_{{\rm{exc}}{\bf{k}}}} = \frac{1}{\sqrt{N_{\rm{mol}}}}\sum\limits_{j}\hat\sigma_{{\rm{exc}}{j}}{e^{i{\bf k}{\bf r}_j}}
 \end{equation}
 where $N_{\rm{mol}}$ is the total number of MeLPPP $\pi$--conjugated segments and ${\bf{k}}$ is the wavevector. Operators $\hat c_{{\rm{exc}}{\bf{k}}}^\dag$ and ${\hat c_{{\rm{exc}}{\bf{k}}}}$ are the creation and annihilation operators of the excitons which obey commutation relation $\left[ {{{\hat c}_{{\rm{exc}}{\bf{k}}}},\hat c_{{\rm{exc}}{\bf{k'}}}^\dag } \right] = {\delta _{{\bf{k}},{\bf{k'}}}}$. Using the operators ${\hat c_{{\rm{exc}}{\bf{k}}}}$ we rewrite the Hamiltonians~(\ref{ExcitonsHamiltonianRealSpace})~and~(\ref{ExcitonsCavityHamiltonianRealSpace}) in the following form
\begin{equation} \label{ExcitonsHamiltonian}
{\hat H_{{\rm{exc}}}} = \sum\limits_{\bf{k}} {\hbar {\omega _{{\rm{exc}}}}\hat c_{{\rm{exc}}{\bf{k}}}^\dag {{\hat c}_{{\rm{exc}}{\bf{k}}}}}
\end{equation}
where ${\omega _{{\rm{exc}}{\bf{k}}}}$ is the eigenfrequency of the exciton with the wavevector ${\bf{k}}$.

\begin{equation} \label{ExcitonsCavityHamiltonian}
{\hat H_{{\rm{exc - cav}}}} = \sum\limits_{\bf{k}} {\hbar {\Omega _{{\rm{R}}{\bf{k}}}}\left( {\hat c_{{\rm{exc}}{\bf{k}}}^\dag {{\hat a}_{{\rm{cav}}{\bf{k}}}} + {{\hat c}_{{\rm{exc}}{\bf{k}}}}\hat a_{{\rm{cav}}{\bf{k}}}^\dag } \right)}
\end{equation}
where ${\Omega _{{\rm{R}}{\bf{k}}}} = \sqrt {{N_{\rm{mol}}}} {\Omega _{{\rm{1R}}{\bf{k}}}}$ is the Rabi frequency of exciton-photon interaction for the excitons and the cavity mode with ${\hbar\bf{k}}$ in-plane momentum. 

The Hamiltonian of vibrons ${\hat H_{{\rm{vib}}}}$ is
\begin{equation} \label{VibronsHamiltonian1}
{\hat H_{{\rm{vib}}}} = \sum\limits_j {\hbar {\omega _{{\rm{vib}}}}\hat b_j^\dag {{\hat b}_j}} \end{equation}
where $\hat b_{j}^\dag$ and ${\hat b_{j}}$ are the creation and annihilation operators of vibrons which obey commutation relation $\left[ {{{\hat b}_j},\hat b_{j'}^\dag } \right] = {\delta _{j,j'}}$.

The Hamiltonian of the interaction between the excitons and vibrons at a single elemental cite of MeLPPP is Fr\"ohlich-type Hamiltonian \cite{Born-QuantChem-2000}
\begin{equation} \label{VibronsHamiltonian1}
{\hat H_{{\rm{exc - vib}}}} = \sum\limits_j {\hbar {g}\hat \sigma _j^\dag {{\hat \sigma }_j}\left( {{{\hat b}_j} + b_j^\dag } \right)} 
\end{equation}
where the $\bf{{g}}$ is the interaction constant between the excitons and vibrons.

Following the same momentum space representation, it is convenient to introduce collective vibron operators according to
\begin{equation} \label{VibronsFourier}
 {\hat b_{{\bf{q}}}} = \frac{1}{\sqrt{N_{\rm{mol}}}}\sum\limits_{j}\hat b_{{j}}{e^{i{\bf q}{\bf r}_j}}
 \end{equation}

Under the transformation (\ref{ExcitonsFourier}) and (\ref{VibronsFourier}) the Hamiltonian of collective vibrons $\hat b_{\bf{k}}$ and its interaction with collective excitons $\hat c_{\rm{exc} \bf{k}}$ takes the form
\begin{equation} \label{VibronsHamiltonian}
{\hat H_{{\rm{vib}}}} = \sum\limits_{\bf{q}} {\hbar {\omega _{{\rm{vib}}}}\hat b_{\bf{q}}^\dag {{\hat b}_{\bf{q}}}}
\end{equation}
\begin{equation} \label{ExcitonsVibronsHamiltonian}
{\hat H_{{\rm{exc - vib}}}} = \sum\limits_{\bf{k}} {\sum\limits_{{\bf{k'}}} {\hbar {g}\hat c_{{\rm{exc}}{\bf{k}}}^\dag {{\hat c}_{{\rm{exc}}{\bf{k'}}}}\left( {\hat b_{ - \left( {{\bf{k}} - {\bf{k'}}} \right)}^\dag  + {{\hat b}_{{\bf{k}} - {\bf{k'}}}}} \right)} }
\end{equation}
where $\hat b_{\bf{q}}^\dag$ and ${\hat b_{\bf{q}}}$ are the creation and annihilation operators of the vibrons which obey commutation relation $\left[ {{{\hat b}_{\bf{q}}},\hat b_{{\bf{q'}}}^\dag } \right] = {\delta _{{\bf{q}},{\bf{q'}}}}$, the vector ${\bf{q}}$ is in-plane wavevector of the vibrons, ${\omega _{{\rm{vib}}}}$ is the eigenfrequency of the vibrons with the in-plane wavevector ${\bf{q}}$. Note that momenta of excitons and vibrons in Eq.~(\ref{ExcitonsVibronsHamiltonian}) fulfils momentum conservation law.

The Hamiltonian of the interaction between excitons and cavity modes ${\hat H_{{\rm{exc - cav}}}}$ is
\begin{equation} \label{ExcitonsCavityHamiltonian}
{\hat H_{{\rm{exc - cav}}}} = \sum\limits_{\bf{k}} {\hbar {\Omega _{{\rm{R}}{\bf{k}}}}\left( {\hat c_{{\rm{exc}}{\bf{k}}}^\dag {{\hat a}_{{\rm{cav}}{\bf{k}}}} + {{\hat c}_{{\rm{exc}}{\bf{k}}}}\hat a_{{\rm{cav}}{\bf{k}}}^\dag } \right)}
\end{equation}
where ${\Omega _{{\rm{R}}{\bf{k}}}}$ is the Rabi frequency of all excitons interacting with the cavity mode specified at the given in-plane momentum ${\bf{\hbar}\bf{k}}$.

The Hamiltoninan of the interaction between the excitons and the external pump field ${\hat H_{{\rm{exc - pump}}}}$ is
\begin{equation} \label{ExcitonsFieldHamiltonian}
{\hat H_{{\rm{exc - pump}}}} = \sum\limits_{\bf{k}} {\hbar {\Omega _{\bf{k}}}\left( t \right)\left( {\hat c_{{\rm{exc}}{\bf{k}}}^\dag {e^{ - i{\omega _\Omega }t}} + {{\hat c}_{{\rm{exc}}{\bf{k}}}}{e^{i{\omega _\Omega }t}}} \right)}
\end{equation}
where ${\Omega _{\bf{k}}}\left( t \right)$ is proportional to the amplitude of the external field and ${\omega _\Omega }$ is the frequency of the external field.

The term in Hamiltoninan describing interaction between the seed electromagnetic field with cavity modes ${\hat H_{{\rm{cav - seed}}}}$ is
\begin{equation} \label{CavityFieldHamiltonian}
{\hat H_{{\rm{cav - seed}}}} = \sum\limits_{\bf{k}} {\hbar {W _{\bf{k}}}\left( t \right)\left( {\hat a_{{\rm{cav}}{\bf{k}}}^\dag {e^{ - i{\omega _W }t}} + {{\hat a}_{{\rm{cav}}{\bf{k}}}}{e^{i{\omega _W }t}}} \right)}
\end{equation}
where ${W _{\bf{k}}}\left( t \right)$ is proportional to the amplitude of the seed field and ${\omega _W }$ its frequency.

Next we diagonalize the light-matter interaction part of the Hamiltonian \cite{ScullyZubairy1997}, namely ${\hat H_{{\rm{cav}}}} + {\hat H_{{\rm{exc}}}} + {\hat H_{{\rm{exc - cav}}}}$. In the new basis of exciton--polariton states, annihilation operators for the lower ${\hat s_{{\rm{low}}{\bf{k}}}}$ and upper ${\hat s_{{\rm{up}}{\bf{k}}}}$ polaritons can be expressed through the standard transformation relations:
\begin{equation} \label{TransformationForLowerPolaritons}
{\hat s_{{\rm{low}}{\bf{k}}}} = {\hat a_{{\rm{cav}}{\bf{k}}}}\cos {\varphi _{\bf{k}}} - {\hat c_{{\rm{exc}}{\bf{k}}}}\sin {\varphi _{\bf{k}}}
\end{equation}
\begin{equation} \label{TransformationForUpperPolaritons}
{\hat s_{{\rm{up}}{\bf{k}}}} = {\hat c_{{\rm{exc}}{\bf{k}}}}\cos {\varphi _{\bf{k}}} + {\hat a_{{\rm{cav}}{\bf{k}}}}\sin {\varphi _{\bf{k}}}
\end{equation}
where
\begin{equation} \label{TransformationAngle}
{\varphi _{\bf{k}}}=
\frac{1}{2}{\rm arctg}\left(       
\frac{2 \Omega_{{\rm R}{\bf k}}}{\omega _{{\rm{exc}}} - \omega _{{\rm{cav}}{\bf{k}}}}
 \right)
\end{equation}
Following the transformations above the full Hamiltonian~(\ref{FullHamiltonian}) takes the form:
\begin{multline}
\hat H = \sum\limits_{\bf{k}} {\hbar {\omega _{{\rm{up}}{\bf{k}}}}\hat s_{{\rm{up}}{\bf{k}}}^\dag {{\hat s}_{{\rm{up}}{\bf{k}}}}}  + \sum\limits_{\bf{k}} {\hbar {\omega _{{\rm{low}}{\bf{k}}}}\hat s_{{\rm{low}}{\bf{k}}}^\dag {{\hat s}_{{\rm{low}}{\bf{k}}}}}  + \sum\limits_{\bf{q}} {\hbar {\omega _{{\rm{vib}}}}\hat b_{\bf{q}}^\dag {{\hat b}_{\bf{q}}}}  + \\
 + \sum\limits_{\bf{k}} {\sum\limits_{{\bf{k'}}} {\hbar {g}\left( {\hat s_{{\rm{up}}{\bf{k}}}^\dag \cos {\varphi _{\bf{k}}} - \hat s_{{\rm{low}}{\bf{k}}}^\dag \sin {\varphi _{\bf{k}}}} \right)\left( {{{\hat s}_{{\rm{up}}{\bf{k'}}}}\cos {\varphi _{{\bf{k'}}}} - {{\hat s}_{{\rm{low}}{\bf{k'}}}}\sin {\varphi _{{\bf{k'}}}}} \right)\left( {\hat b_{ - \left( {{\bf{k}} - {\bf{k'}}} \right)}^\dag  + {{\hat b}_{{\bf{k}} - {\bf{k'}}}}} \right)} }  + \\
 + \sum\limits_{\bf{k}} {\hbar {\Omega _{\bf{k}}}\left( t \right)\left( {\left( {\hat s_{{\rm{up}}{\bf{k}}}^\dag \cos {\varphi _{\bf{k}}} - \hat s_{{\rm{low}}{\bf{k}}}^\dag \sin {\varphi _{\bf{k}}}} \right){e^{ - i{\omega _\Omega }t}} + \left( {{{\hat s}_{{\rm{up}}{\bf{k'}}}}\cos {\varphi _{{\bf{k'}}}} - {{\hat s}_{{\rm{low}}{\bf{k'}}}}\sin {\varphi _{{\bf{k'}}}}} \right){e^{i{\omega _\Omega }t}}} \right)}  + \\
 + \sum\limits_{\bf{k}} {\hbar {W_{\bf{k}}}\left( t \right)\left( {\left( {\hat s_{{\rm{up}}{\bf{k}}}^\dag \cos {\varphi _{\bf{k}}} - \hat s_{{\rm{low}}{\bf{k}}}^\dag \sin {\varphi _{\bf{k}}}} \right){e^{ - i{\omega _W}t}} + \left( {{{\hat s}_{{\rm{up}}{\bf{k'}}}}\cos {\varphi _{{\bf{k'}}}} - {{\hat s}_{{\rm{low}}{\bf{k'}}}}\sin {\varphi _{{\bf{k'}}}}} \right){e^{i{\omega _W}t}}} \right)}
\end{multline}
where the new eigenfrequencies are:

\begin{equation} \label{FrequenciesOfLowerPolaritons}
{\omega _{{\rm{low}}{\bf{k}}}} = {{\left( {{\omega _{{\rm{exc}}}} + {\omega _{{\rm{cav}}{\bf{k}}}}} \right)} \mathord{\left/
 {\vphantom {{\left( {{\omega _{{\rm{exc}}}} + {\omega _{{\rm{cav}}{\bf{k}}}}} \right)} 2}} \right.
 \kern-\nulldelimiterspace} 2} - \sqrt {{{{{\left( {{\omega _{{\rm{exc}}}} - {\omega _{{\rm{cav}}{\bf{k}}}}} \right)}^2}} \mathord{\left/
 {\vphantom {{{{\left( {{\omega _{{\rm{exc}}}} - {\omega _{{\rm{cav}}{\bf{k}}}}} \right)}^2}} 4}} \right.
 \kern-\nulldelimiterspace} 4} + \Omega _{{\rm{R}}{\bf{k}}}^2}
\end{equation}

\begin{equation} \label{FrequenciesOfUpperPolaritons}
{\omega _{{\rm{up}}{\bf{k}}}} = {{\left( {{\omega _{{\rm{exc}}}} + {\omega _{{\rm{cav}}{\bf{k}}}}} \right)} \mathord{\left/
 {\vphantom {{\left( {{\omega _{{\rm{exc}}}} + {\omega _{{\rm{cav}}{\bf{k}}}}} \right)} 2}} \right.
 \kern-\nulldelimiterspace} 2} + \sqrt {{{{{\left( {{\omega _{{\rm{exc}}}} - {\omega _{{\rm{cav}}{\bf{k}}}}} \right)}^2}} \mathord{\left/
 {\vphantom {{{{\left( {{\omega _{{\rm{exc}}}} - {\omega _{{\rm{cav}}{\bf{k}}}}} \right)}^2}} 4}} \right.
 \kern-\nulldelimiterspace} 4} + \Omega _{{\rm{R}}{\bf{k}}}^2}
\end{equation}

In turn, lower and upper polaritons as well as vibrons interact with the environment that inevitably leads to the relaxation processes. We consider the relaxation processes by means of Lindblad superoperators \cite{ScullyZubairy1997,Carmichael1991} acting on the general density matrix operator $\hat \rho$ which ultimately describes the entire system including all the polaritons and vibrons. The Lindblad superoperators for the lower, upper polaritons and vibrons are read explicitly

\begin{equation} \label{LindbladDissipatonLowerPolaritons}
{L_{{\rm{low}}}}\left( {\hat \rho } \right) = \sum\limits_{\bf{k}} {\frac{{{\gamma _{{\rm{low}}{\bf{k}}}}}}{2}\left( {2{{\hat s}_{{\rm{low}}{\bf{k}}}}\hat \rho \hat s_{{\rm{low}}{\bf{k}}}^\dag  - \hat s_{{\rm{low}}{\bf{k}}}^\dag {{\hat s}_{{\rm{low}}{\bf{k}}}}\hat \rho  - \hat \rho \hat s_{{\rm{low}}{\bf{k}}}^\dag {{\hat s}_{{\rm{low}}{\bf{k}}}}} \right)}
\end{equation}

\begin{equation} \label{LindbladDissipatonUpperPolaritons}
{L_{{\rm{up}}}}\left( {\hat \rho } \right) = \sum\limits_{\bf{k}} {\frac{{{\gamma _{{\rm{up}}{\bf{k}}}}}}{2}\left( {2{{\hat s}_{{\rm{up}}{\bf{k}}}}\hat \rho \hat s_{{\rm{up}}{\bf{k}}}^\dag  - \hat s_{{\rm{up}}{\bf{k}}}^\dag {{\hat s}_{{\rm{up}}{\bf{k}}}}\hat \rho  - \hat \rho \hat s_{{\rm{up}}{\bf{k}}}^\dag {{\hat s}_{{\rm{up}}{\bf{k}}}}} \right)}
\end{equation}

\begin{equation} \label{LindbladDissipatonVibrons}
{L_{{\rm{vib}}}}\left( {\hat \rho } \right) = \sum\limits_{\bf{q}} {\frac{{{\gamma _{{\rm{vib}}}}}}{2}\left( {1 + n_{\rm{vib}}^{{\rm{th}}}} \right)\left( {2{{\hat b}_{\bf{q}}}\hat \rho \hat b_{\bf{q}}^\dag  - \hat b_{\bf{q}}^\dag {{\hat b}_{\bf{q}}}\hat \rho  - \hat \rho \hat b_{\bf{q}}^\dag {{\hat b}_{\bf{q}}}} \right)}
 + \sum\limits_{\bf{q}} {\frac{{{\gamma _{{\rm{vib}}}}}}{2}n_{\rm{vib}}^{{\rm{th}}}\left( {2\hat b_{\bf{q}}^\dag \hat \rho {{\hat b}_{\bf{q}}} - {{\hat b}_{\bf{q}}}\hat b_{\bf{q}}^\dag \hat \rho  - \hat \rho {{\hat b}_{\bf{q}}}\hat b_{\bf{q}}^\dag } \right)}
\end{equation}
where ${\gamma _{{\rm{low}}{\bf{k}}}}$, ${\gamma _{{\rm{up}}{\bf{k}}}}$ and ${\gamma _{{\rm{vib}}}}$ are the dissipation rates of lower, upper polaritons and the vibrons respectively, and $n_{\rm{vib}}^{{\rm{th}}} = {1 \mathord{\left/
 {\vphantom {1 {\left( {\exp \left( {{{\hbar {\omega _{{\rm{vib}}}}} \mathord{\left/
 {\vphantom {{\hbar {\omega _{{\rm{vib}}}}} T}} \right.
 \kern-\nulldelimiterspace} T}} \right) - 1} \right)}}} \right.
 \kern-\nulldelimiterspace} {\left( {\exp \left( {{{\hbar {\omega _{{\rm{vib}}}}} \mathord{\left/
 {\vphantom {{\hbar {\omega _{{\rm{vib}}{\bf{q}}}}} T}} \right.
 \kern-\nulldelimiterspace} T}} \right) - 1} \right)}}$ is the thermal distribution of vibrons at an effective lattice temperature $T$.

Thermalization of the lower polaritons is described by the following Lindblad superoperator:

\begin{equation} \label{LindbladThermalizationLowerPolaritons}
{\hat L_{{\rm{th}}}}\left( {\hat \rho } \right) = \sum\limits_{{{\bf{k}}_1},{{\bf{k}}_2}} {\frac{{\gamma _{{\rm{low}}}^{{{\bf{k}}_1}{{\bf{k}}_2}}}}{2}\left( {2{{\hat s}_{{\rm{low}}{{\bf{k}}_2}}}\hat s_{{\rm{low}}{{\bf{k}}_1}}^\dag \hat \rho {{\hat s}_{{\rm{low}}{{\bf{k}}_1}}}\hat s_{{\rm{low}}{{\bf{k}}_2}}^\dag  - {{\hat s}_{{\rm{low}}{{\bf{k}}_1}}}\hat s_{{\rm{low}}{{\bf{k}}_2}}^\dag {{\hat s}_{{\rm{low}}{{\bf{k}}_2}}}\hat s_{{\rm{low}}{{\bf{k}}_1}}^\dag \hat \rho  - \hat \rho {{\hat s}_{{\rm{low}}{{\bf{k}}_1}}}\hat s_{{\rm{low}}{{\bf{k}}_2}}^\dag {{\hat s}_{{\rm{low}}{{\bf{k}}_2}}}\hat s_{{\rm{low}}{{\bf{k}}_1}}^\dag } \right)} 
\end{equation}
here $\gamma _{{\rm{low}}}^{{{\bf{k}}_1}{{\bf{k}}_2}}$ is the rate of energy flow from the lower polaritons having in-plane wavevector ${{\bf{k}}_2}$ towards ones with  wavevector ${{\bf{k}}_1}$. According to  Kubo-Martin-Schwinger relation, the rate $\gamma _{{\rm{low}}}^{{{\bf{k}}_1}{{\bf{k}}_2}}$ is defined as

\begin{equation} \label{KMShLowerThermalizationRates}
\gamma _{{\rm{low}}}^{{{\bf{k}}_1}{{\bf{k}}_2}} = \gamma _{{\rm{low}}}^{{{\bf{k}}_2}{{\bf{k}}_1}}\exp \left( {\frac{{{\omega _{{\rm{low}}{{\bf{k}}_2}}} - {\omega _{{\rm{low}}{{\bf{k}}_1}}}}}{T}} \right)
\end{equation}
where $T$ is the effective MeLPPP temperature.

The resulting master equation is
\begin{equation} \label{MasterEquation}
{\frac{ d\hat \rho}{dt}} = {\frac{ i}{\hbar}}{\left[ {\hat \rho}, {\hat H} \right]}
+ {\hat L_{{\rm{up}}}}\left( {\hat \rho } \right)
+ {\hat L_{{\rm{low}}}}\left( {\hat \rho } \right)
+ {\hat L_{{\rm{vib}}}}\left( {\hat \rho } \right)
+ {\hat L_{{\rm{th}}}}\left( {\hat \rho } \right)
\end{equation}

Further on we develop a mean-field theory \cite{Carmichael1991} for the system described above. The mean-field theory describes dynamics of the system through evolution of amplitudes of upper polaritons, ${s_{{\rm{up}}{\bf{k}}}} = \left\langle {{{\hat s}_{{\rm{up}}{\bf{k}}}}} \right\rangle $, lower polaritons, ${s_{{\rm{low}}{\bf{k}}}} = \left\langle {{{\hat s}_{{\rm{low}}{\bf{k}}}}} \right\rangle $, and vibrons, ${b_{\bf{q}}} = \left\langle \hat b_{\bf{q}} \right\rangle $. To derive these equations we use the following definition $\left\langle {\dot {\hat A}} \right\rangle  = {\rm{Tr}}\left( {\dot{ \hat \rho} \hat A} \right)$ and the master equation~(\ref{MasterEquation}). The resulting mean-field equations take form
\begin{equation} \label{AmplitudeUpperPolaritons}
\frac{{d{s_{{\rm{up}}{\bf{k}}}}}}{{dt}} =  - i{\omega _{{\rm{up}}{\bf{k}}}}{s_{{\rm{up}}{\bf{k}}}} - \frac{1}{2}{\gamma _{{\rm{up}}{\bf{k}}}}{s_{{\rm{up}}{\bf{k}}}} + i\sum\limits_{{\bf{k'}}} {{g}\cos {\varphi _{\bf{k}}}\sin {\varphi _{{\bf{k'}}}}{s_{{\rm{low}}{\bf{k'}}}}{b_{{\bf{k}} - {\bf{k'}}}}}  - i{\Omega _{\bf{k}}}\left( t \right)\cos {\varphi _{\bf{k}}}{e^{ - i{\omega _\Omega }t}}
\end{equation}
\begin{multline} \label{AmplitudeLowerPolaritons}
\frac{{d{s_{{\rm{low}}{\bf{k}}}}}}{{dt}} =  - i{\omega _{{\rm{low}}{\bf{k}}}}{s_{{\rm{low}}{\bf{k}}}} - \frac{1}{2}{\gamma _{{\rm{low}}{\bf{k}}}}{s_{{\rm{low}}{\bf{k}}}} + i\sum\limits_{{\bf{k'}}} {{g}\cos {\varphi _{{\bf{k'}}}}\sin {\varphi _{\bf{k}}}{s_{{\rm{up}}{\bf{k'}}}}b_{{\bf{k'}} - {\bf{k}}}^*}
+ 
\\
+ i\sum\limits_{{\bf{k'}}} {{g}\sin {\varphi _{{\bf{k'}}}}\sin {\varphi _{\bf{k}}}{s_{{\rm{low}}{\bf{k'}}}}(b_{{\bf{k'}} - {\bf{k}}}^*+b_{{\bf{k}}-{\bf{k'}}})}- i{\Omega _{\bf{k}}}\left( t \right)\sin {\varphi _{\bf{k}}}{e^{ - i{\omega _\Omega }t}} - i{W_{\bf{k}}}\left( t \right)\cos {\varphi _{\bf{k}}}{e^{ - i{\omega _W}t}} +
\\
+\frac{1}{2}\sum\limits_{{\bf{k'}}} {\left( {\gamma _{{\rm{low}}}^{{\bf{kk'}}}{{\left| {{s_{{\rm{low}}{\bf{k'}}}}} \right|}^2} - \gamma _{{\rm{low}}}^{{\bf{k'k}}}\left( {1 + {{\left| {{s_{{\rm{low}}{\bf{k'}}}}} \right|}^2}} \right)} \right)} {s_{{\rm{low}}{\bf{k}}}}
\end{multline}
\begin{multline} \label{AmplitudeVibrons}
\frac{d{b_{\bf{q}}}}{dt} =  - i{\omega _{{\rm{vib}}}}{b_{\bf{q}}} - \frac{1}{2}{\gamma _{{\rm{vib}}}}{b_{\bf{q}}}+i\sum\limits_{{\bf{k'}}} {g\cos {\varphi _{{\bf{k'}}}}\sin {\varphi _{{\bf{k'}} - {\bf{q}}}}{s_{{\rm{up}}{\bf{k'}}}}s_{{\rm{low}}{\bf{k'}} - {\bf{q}}}^*}+
\\
+i\sum\limits_{{\bf{k'}}} {g\sin {\varphi _{{\bf{k'}}}}\sin {\varphi _{{\bf{k'}} - {\bf{q}}}}{s_{{\rm{low}}{\bf{k'}}}}s_{{\rm{low}}{\bf{k'}} - {\bf{q}}}^*}
\end{multline}
Here we neglect non-resonant terms, that reasonably simplifies Eqs.~(\ref{AmplitudeUpperPolaritons})-(\ref{AmplitudeVibrons}).

To discretize the system we map its momenta states onto the energy space, moving from the continuum of wave vectors to a discrete set of frequencies. Note that since Eq.~(\ref{AmplitudeLowerPolaritons}) contains spontaneous term, ${ {1 + {{\left| {{s_{{\rm{low}}{\bf{k'}}}}} \right|}^2}} }$, one should take into account the density of states. We consider a discrete finite set of frequencies (${\omega _0}$, ${\omega _1}$, ${\omega _2}$, ..., ${\omega _N}$) with a sampling rate equal to $\delta \omega $, that is ${\omega _{j + 1}} - {\omega _j} = \delta \omega $, where $N$ means the total number of sampled frequencies. For the sake of simplicity we introduce the ground state frequency as ${\omega _0} = {\omega _{{\rm{low}}\left( {{\bf{k}} = 0} \right)}}$ and determine the amplitude of the descretized lower polariton ${s_{{\rm{low}}{\omega _j}}}$ with the eigenfrequency ${\omega _j}$ as follows
\begin{equation} \label{LowerPolaritonsDescrite}
{s_{{\rm{low}}{\omega _j}}} = \sum\limits_{{\bf{k}},\,\,{\omega _j} < {\omega _{{\rm{low}}{\bf{k}}}} < {\omega _j} + \delta \omega } {{s_{{\rm{low}}{\bf{k}}}}}
\end{equation}

To calculate the number of states within the frequency range ${\omega _j} \le \omega  \le {\omega _j} + \delta \omega $ we approximate lower polariton dispersion with a quadratic dependence ${\omega _{{\rm{low}}{\bf{k}}}} = {\omega _{{\rm{low}}\left( {{\bf{k}} = 0} \right)}} + \alpha_{\rm{cav}} {{\bf{k}}^2}$ so that it appears explicitly in the following form:
\begin{equation} \label{NumberOfStates}
{D_{{\omega _j}}} = \sum\limits_{{\bf{k}},\,\,{\omega _j} < {\omega _{{\rm{low}}{\bf{k}}}} < {\omega _j} + \delta \omega } 1  \approx \left( {Ln_{{\rm{MeLPPP}}}^{1/3}} \right)\frac{S}{{{{\left( {2\pi } \right)}^2}}}\int\limits_{{\omega _j} < {\omega _{{\rm{low}}{\bf{k}}}} < {\omega _j} + \delta \omega } {{d^2}{\bf{k}}}  = \frac{S}{{{{\left( {2\pi } \right)}^2}}}\left( {Ln_{{\rm{MeLPPP}}}^{1/3}} \right)\frac{\pi }{\alpha }\delta \omega
\end{equation}
where $S$ is the area illuminated by the pump, $L$ is the thickness of MeLPPP layer, ${n_{{\rm{MeLPPP}}}}$ is the concentration of MeLPPP. Unity in Eq.~(\ref{NumberOfStates}) denotes the one state with fixed wavevector $\bf{k}$. Note that the expression for ${D_{{\omega _j}}}$ is frequency independent in the case under consideration. If we substitute characteristic values of $S = {300}\,\,{\rm{\mu }}{{\rm{m}}^2}$, $L = 0.035\,\,{\rm{\mu m}}$, ${n_{{\rm{MeLPPP}}}} \approx 10^{6}\,\,{\rm{\mu m}}^{-3}$  and $\alpha_{\rm{cav}}  = 2.2 \cdot {10^{ - 3}}\,\,{\rm{eV}} \cdot {\rm{\mu }}{{\rm{m}}^2}$, then we obtain ${D_{{\omega _j}}} \approx {10^4} \cdot \delta \omega$ where $\delta \omega $ is measured in ${\rm{eV}}$. The occupancy of the discrete modes at the lower polariton branch $n_{{\rm low}j}$  takes the form
\begin{equation} \label{occupancyLowDescriteDefenition}
{n_{{\rm{low}}{\omega _j}}} = \sum\limits_{{\bf{k}},\,\,{\omega _j} < {\omega _{{\rm{low}}{\bf{k}}}} < {\omega _j} + \delta \omega } {|{s_{{\rm{low}}{\bf{k}}}}{|^2}}  = \frac{{|{s_{{\rm{low}}{\omega _j}}}{|^2}}}{{{D_{{\omega _j}}}}}
\end{equation}

We determine the descritized amplitude of the vibrons ${b_{{\rm{vib}}j}}$ according to
\begin{equation} \label{VibronsDescrete}
{b_{{\rm{vib}}j}} = \sum\limits_{{\bf{q}},\,\,{\omega _j} < {\omega _{{\rm{low}}\left( {{{\bf{k}}_{{\rm{ex}}}} - {\bf{q}}} \right)}} < {\omega _j} + \delta \omega } {{b_{{\bf{q}}}}}
\end{equation}
The summation~({\ref{VibronsDescrete}}) takes only over the vibrons participating energy transfer from the upper and lower polariton modes with the wavevector ${{\bf{k}}_{{\rm{ex}}}}$ towards the discretized lower polariton mode with frequency ${\omega _j}$. as 
\begin{equation} \label{ExternalField}
{\Omega _{\bf{k}}}\left( t \right) = 0,\,\,{\rm{for}}\,\,{\bf{k}} \ne {{\bf{k}}_{{\rm{ex}}}}
\end{equation}

The process of energy transfer from the higher-lying states to the bottom of lower polariton branch can be formally represented as a set of elementary 3-body interactions, where a vibron with a wavevector ${\bf{q}}$ and a lower polariton with a quasimomentum ${{\bf{k}}_{{\rm{ex}}}} - {\bf{q}}$ are created following annihilation of polariton with in-plane momentum ${{\hbar\bf{k}}_{{\rm{ex}}}}$. 
Using momentum-to-energy mapping procedure above~(\ref{VibronsDescrete})-(\ref{ExternalField}), we end up with the descritized version of the Eqs.~(\ref{AmplitudeUpperPolaritons})-(\ref{AmplitudeVibrons}) in frequency space.
\begin{equation} \label{AmplitudeUpperPolaritonsDescrete}
\frac{{d{s_{{\rm{up}}{{\bf{k}}_{{\rm{ex}}}}}}}}{{dt}} =  - i{\omega _{{\rm{up}}{{\bf{k}}_{{\rm{ex}}}}}}{s_{{\rm{up}}{{\bf{k}}_{{\rm{ex}}}}}} - \frac{1}{2}{\gamma _{{\rm{up}}{{\bf{k}}_{{\rm{ex}}}}}}{s_{{\rm{up}}{{\bf{k}}_{{\rm{ex}}}}}} +ig\sum\limits_{j = 0}^N {\cos {\varphi _{{{\bf{k}}_{{\rm{ex}}}}}}\frac{\sin {\varphi _j}{s_{{\rm{low}}{\omega _j}}}{b_{{\rm vib}j}}}{D_{\omega_j}} -
i{\Omega _{{{\bf{k}}_{{\rm{ex}}}}}}\left( t \right)\cos {\varphi _{{{\bf{k}}_{{\rm{ex}}}}}}{e^{ - i{\omega _\Omega }t}}}
\end{equation}
\begin{equation} \label{AmplitudeLowerPolaritonsDescreteEx}
\frac{{d{s_{{\rm{low}}{{\bf{k}}_{{\rm{ex}}}}}}}}{{dt}} =  - i{\omega _{{\rm{low}}{{\bf{k}}_{{\rm{ex}}}}}}{s_{{\rm{low}}{{\bf{k}}_{{\rm{ex}}}}}} - \frac{1}{2}{\gamma _{{\rm{low}}{{\bf{k}}_{{\rm{ex}}}}}}{s_{{\rm{up}}{{\bf{k}}_{{\rm{ex}}}}}} + ig\sum\limits_{j = 0}^N {\sin {\varphi _{{{\bf{k}}_{{\rm{ex}}}}}}\frac{\sin {\varphi _j}{s_{{\rm{low}}{\omega _j}}}{b_{{\rm vib}j}}}{D_{\omega_j}} - i{\Omega _{{{\bf{k}}_{{\rm{ex}}}}}}\left( t \right)\sin {\varphi _{{{\bf{k}}_{{\rm{ex}}}}}}{e^{ - i{\omega _\Omega }t}}}
\end{equation}
\begin{multline} \label{AmplitudeLowerPolaritonsDescrete}
\frac{{d{s_{{\rm{low}}{\omega _j}}}}}{{dt}} =
- i{\omega _j}{s_{{\rm{low}}{\omega _j}}}
- \frac{1}{2}{\gamma _{{\rm{low}}{\omega _j}}}{s_{{\rm{low}}{\omega _j}}}
+ ig\cos {\varphi _{{{\bf{k}}_{{\rm{ex}}}}}}\sin {\varphi _j}{s_{{\rm{up}}{{\bf{k}}_{{\rm{ex}}}}}}b_{{\rm vib}j}^*
+ ig\sin {\varphi _{{{\bf{k}}_{{\rm{ex}}}}}}\sin {\varphi _j}{s_{{\rm{low}}{{\bf{k}}_{{\rm{ex}}}}}}b_{{\rm vib}j}^*-
\\- i{W_{{\omega _j}}}\left( t \right)\cos {\varphi _j}{e^{ - i{\omega _W}t}}
+ \frac{1}{2}\sum\limits_{m = 0}^N {\left( {\gamma _{{\rm{low}}}^{{\omega _j}{\omega _m}}\frac{{\left| {{s_{{\rm{low}}{\omega _m}}}} \right|}^2}{D_{\omega_m}} - \gamma _{{\rm{low}}}^{{\omega _m}{\omega _j}}\left( {D_{\omega_m} + \frac{{\left| {{s_{{\rm{low}}{\omega _m}}}} \right|}^2}{D_{\omega_m}}} \right)} \right)} {s_{{\rm{low}}{\omega _j}}}
\end{multline}
\begin{equation} \label{AmplitudeVibronsDescrete}
\frac{{d{b_{{\rm vib}j}}}}{{dt}} =  - i{\omega _{{\rm{vib}}}}{b_{{\rm vib}j}} - \frac{1}{2}{\gamma _{{\rm{vib}}}}{b_{{\rm vib}j}}
+ ig\cos {\varphi _{{{\bf{k}}_{{\rm{ex}}}}}}\sin {\varphi _j}{s_{{\rm{up}}{{\bf{k}}_{{\rm{ex}}}}}}s_{{\rm{low}}{\omega _j}}^*
+ ig\sin {\varphi _{{{\bf{k}}_{{\rm{ex}}}}}}\sin {\varphi _j}{s_{{\rm{low}}{{\bf{k}}_{{\rm{ex}}}}}}s_{{\rm{low}}{\omega _j}}^*
\end{equation}
where the rates, frequencies and coupling parameters are reformulated accordingly: 
${\gamma _{{\rm{low}}{\omega _j}}} = {\gamma _{{\rm{low}}{\bf{k}}}}$ and $\sin {\varphi _{{\omega _j}}} = \sin {\varphi _{\bf{k}}}$ with ${\bf{k}}$ such that ${\omega _{{\rm{low}}{\bf{k}}}} = {\omega _j}$;  $\gamma _{{\rm{low}}}^{{\omega _j}{\omega _m}} = \gamma _{{\rm{low}}}^{{\bf{kk'}}}$ with ${\bf{k}}$ and ${\bf{k'}}$ such that ${\omega _{{\rm{low}}{\bf{k}}}} = {\omega _j}$ and ${\omega _{{\rm{low}}{\bf{k'}}}} = {\omega _m}$. Here we assume the quantities ${\gamma _{{\rm{low}}{\bf{k}}}}$, $\sin {\varphi _{\bf{k}}}$ and $\gamma _{{\rm{low}}}^{{\bf{kk'}}}$ are changed slightly on the discretization scale $\delta \omega $ which is a reasonable assumption unless one applies $\delta \omega \sim T$. Note the new thermalization rates $\gamma _{{\rm{low}}}^{{\omega _j}{\omega _m}}$ meet the same Kubo-Martin-Schwinger relation
\begin{equation} \label{KMShLowerThermalizationRatesDescrete}
\gamma _{{\rm{low}}}^{{\omega _j}{\omega _m}} = \gamma _{{\rm{low}}}^{{\omega _m}{\omega _j}}\exp \left( {\frac{{{\omega _m} - {\omega _j}}}{T}} \right)
\end{equation}

According to the mapping procedure, we use the following relations in the Eqs. (\ref{AmplitudeUpperPolaritonsDescrete})-(\ref{AmplitudeVibronsDescrete})   

\begin{equation}
\sum\limits_{{\bf{k'}}} {\sin {\varphi _{{\bf{k'}}}}{s_{{\rm{low}}{\bf{k'}}}}{b_{{\bf{k}} - {\bf{k'}}}}} = \sum\limits_{j}{\frac{\sin\varphi_j s_{{\rm low}\omega_j}b_{{\rm{vib}}j}}{D_{\omega_j}}}
\end{equation}

\begin{equation} \label{ProbeDescrete}
{W_{{\rm{low}}{\omega _j}}}\left( t \right) = \sum\limits_{{\bf{k'}},\,\,{\omega _j} < {\omega _{{\rm{low}}{\bf{k'}}}} < {\omega _j} + \delta \omega } {{W_{{\rm{low}}{\bf{k'}}}}\left( t \right)} 
\end{equation}

\section{Numerical simulations}

\begin{figure}[!h]
\centering
\includegraphics[width=0.5\textwidth]{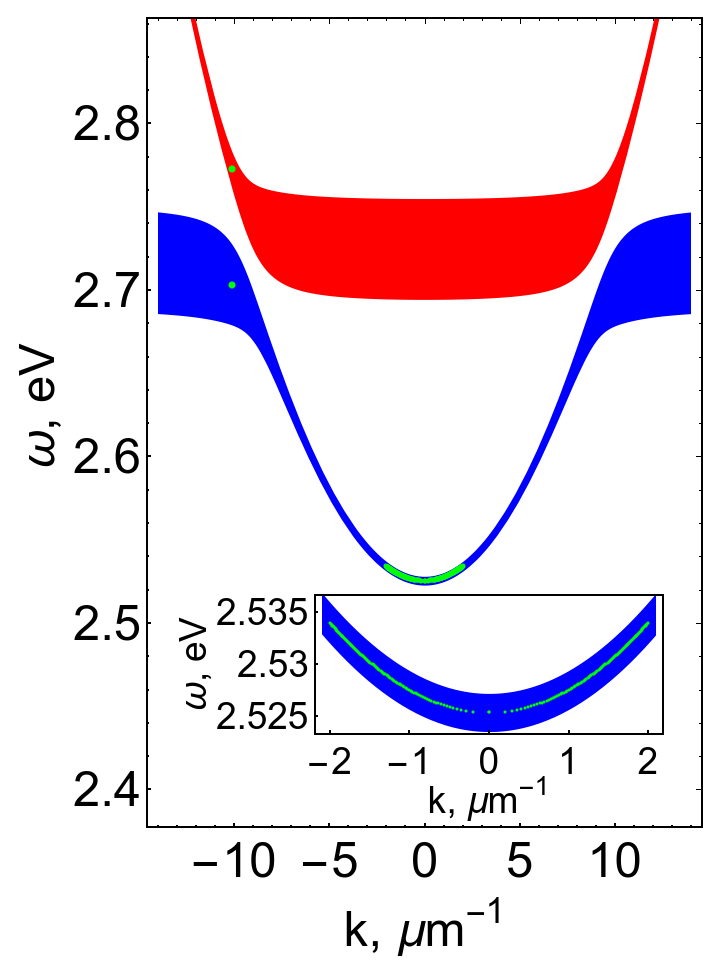}
\caption{The dispersion of the lower (red) and upper (blue) polariton branches. The width of the lines along the frequency axes indicates the FWHM of the polaritons, which is corresponds to the dissipation rate. The descrete modes are shown as green dots.}
\label{fig:Dispersion}
\end{figure}

We generate the system of differential equations based on Eqs. (\ref{AmplitudeUpperPolaritonsDescrete})-(\ref{AmplitudeVibronsDescrete}) consisting of the modes in upper and lower polariton branches at ${{\bf{k}}_{{\rm{ex}}}}$, the most intense vibron mode with energy equals to $\omega_{{\rm vib}}=0.199$~$\rm eV$ and the set of equidistant $N=120$ modes in frequency space that fill the bottom of lower polariton dispersion within the momenta range of [$|k_{\rm min}|=0$~$\rm \mu m^{-1} , |k_{\rm max}|=2$~$\rm \mu m^{-1}$]. All the modes are shown in Fig.\ref{fig:Dispersion} as green dots on top of the polariton dispersion relations. Modes at the bottom of lower polariton branch are shown in the inset of Fig.\ref{fig:Dispersion}. 

To calculate polariton dispersion relations we use standard parabolic dispersion of the cavity $\omega_{\rm{cav}\bf{k}}=\omega_{\rm{cav}0}+\alpha_{\rm{cav}}\cdot{\bf k}^2$, where the cut-off frequency at ${\bf k}=0$ is equal to $\omega_{\rm{cav}0} = 2.53$~eV and the curvature parameter $\alpha_{\rm{cav}}=2.2\cdot10^{-3}$~$\rm{eV}\rm{\mu m}^2$ is given from experimental data. In turn, Frenkel excitons are well-known as dispersionless particles, therefore it is reasonable assuming the constant energy of $\omega_{\rm{exc}} = 2.72$~eV. Similarly, the coupling constant between exciton and cavity modes is fixed at $\Omega_{{\rm R}{\bf k}} = 0.035$~$\rm eV$ regardless in-plane quasimomentum $\bf{k}$. The photon decay and exciton dephasing rates are assumed to be $\bf{k}$-independent values $\gamma_{\rm{cav}\bf{k}}=2\cdot10^{-3}$~$\rm{eV}$ and $\gamma_{\rm{exc}}=0.06$~$\rm{eV}$ respectively. 

Finally, we use Eqs.~(\ref{FrequenciesOfLowerPolaritons}),(\ref{FrequenciesOfUpperPolaritons}) to plot lower and upper polariton branches in Fig.\ref{fig:Dispersion}, where the linewidth for both dependencies correspond to dissipation rate of the lower $\gamma_{{\rm low}{\bf k}}=\gamma_{{\rm cav}{\bf k}}\cos^2\phi_{\bf k}+\gamma_{{\rm exc}}\sin^2\phi_{\bf k}$ and upper $\gamma_{{\rm up}{\bf k}}=\gamma_{{\rm cav}{\bf k}}\sin^2\phi_{\bf k}+\gamma_{{\rm exc}}\cos^2\phi_{\bf k}$ polaritons respectively.

We excite our system by pump electromagnetic field which couples exciton states exactly at their resonance  $\omega_{\Omega}=\omega_{\rm{exc}}=2.72$eV. The incident angle of the pump beam is $\theta=45^{\rm o}$ that corresponds to in-plane momentum ${\hbar\bf k}=10$~$\rm \mu m^{-1}$. The pump pulse duration of $150$~$\rm fs$ corresponds to $t_\Omega=2 \pi \cdot 30$~$\rm eV^{-1}$. The pump frequency fulfils the resonant vibron condition $\omega_{\Omega}=\omega_{{\rm low}0}+\omega_{{\rm vib}}$. The vibron resonance frequency is given from experimental data $\omega_{{\rm vib}{\bf q}}=0.199$~$\rm eV$, while exciton-vibron interaction constant ${\bf{g}}$ is left to be variable parameter during numerical simulations. We get the best fit results with ${g}=0.45\cdot10^{-3}$~$\rm eV$. The value agrees well with the experimental one assessed independently using a cross-section of Raman scattering \cite{Somitsch-SyntMet-2003} and theoretical approach developed in Ref.\cite{Shishkov-PRL-2019}.

 To study dynamics of the polariton condensation and associated nonlinear emission from the organic microcavity we set thermalization rate for the lower polaritons at the level of $\gamma _{{\rm{low}}}^{{{\bf{k}}_1}{{\bf{k}}_2}}=0.7 \cdot 10^{-7}$~$\rm eV$ (it corresponds to the best fit result) for $\omega_{{\rm low}{\bf k_1}}<\omega_{{\rm low}{\bf k_2}}$ and use thermal energy $\textit{kT} = 0.025$~$\rm eV$ corresponding room temperature 300K. The vibron dissipation rate is estimated from Raman spectra of MeLPPP
$\gamma_{{\rm vib}}=0.0025$~eV. Following experimental conditions we choose $|k|<0.2$~$\rm \mu m^{-1}$ and $\omega_{W}=\omega_{{\rm low}0}$ for the seed. The seed pulse duration of $ 200$~$\rm fs$ corresponds to $t_\Omega=2\pi \cdot 40$~$\rm eV^{-1}$. Finally, we apply the following initial conditions to the system: $s_{{\rm up}{\bf k}_{\rm ex}}(0)=0$, $s_{{\rm low}{\omega}_j}(0)=0$ for all $j$,
$b_{j}(0)=\sqrt{n^{{\rm th}}_{\rm{vib}}}$ for all $j$.

\begin{figure}[h!]
    \centering
    \includegraphics[width=0.75\textwidth]{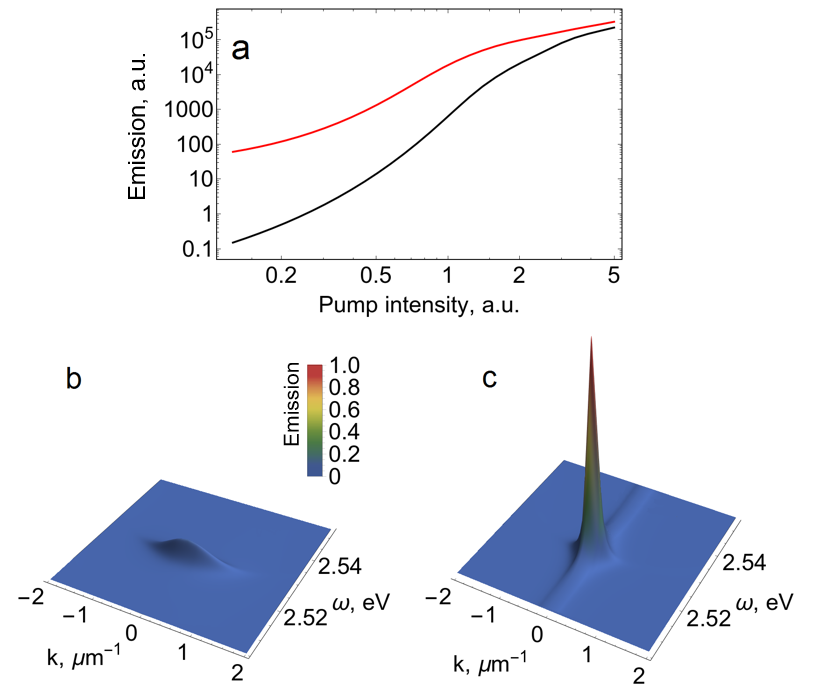}
    \caption{a - Emission of the seeded (red) and spontaneously formed (black condensates integrated over the polariton state within $\pm 0.2$ $\mu m^{-1}$ momentum range as a function of pump intensity $ P$. b and c - Time-integrated energy, momentum \textit{(E,k)}-distribution of polariton population along the lower branch for the spontaneously formed and seeded polariton condensates respectively. Population distributions are simulated at $ P=2P_{th}$}
    \label{fig:figs2}
\end{figure}

We simulate integrated emission as a function of dimensionless pump intensity $P$ for the seeded (Fig.\ref{fig:figs2}a, red) and spontaneously formed (Fig.\ref{fig:figs2}a, black) condensates. The integration takes place within $\pm 0.2$ $\mu m^{-1}$ according to the experimental configurations. One can see remarkable agreement with the experimental curves represented in Fig.1b of the main text. Importantly, our model adequately describes polariton population distribution in \textit{(E,k)}-space. Fig.\ref{fig:figs2}b and c demonstrate spontaneously formed and seeded condensates respectively at the incident pump fluence of $P\sim 2P_{th}$. Here we use the seed pulse carrying 1000 photons which resonantly inject polaritons at the ground state. As we form macroscopic occupation of the ground state at the very beginning stage of polariton condensation, it boosts relaxation towards the ground state through the vibron-assisted bosonic stimulation. One can see the predominant population into the preoccupied states via sharp distribution of the polariton population along the lower branch. Both simulations accurately describe experimental \textit{(E,k)} distributions from Fig.1b,c of the main manuscript. In fact, it manifests our microscopic theory is adequate describing physical mechanisms underlying vibron-mediated polariton condensation in the organic system at resonant excitation and seeding.  
\clearpage

\begin{figure}[h!]
\centering
\includegraphics[width=0.5\textwidth]{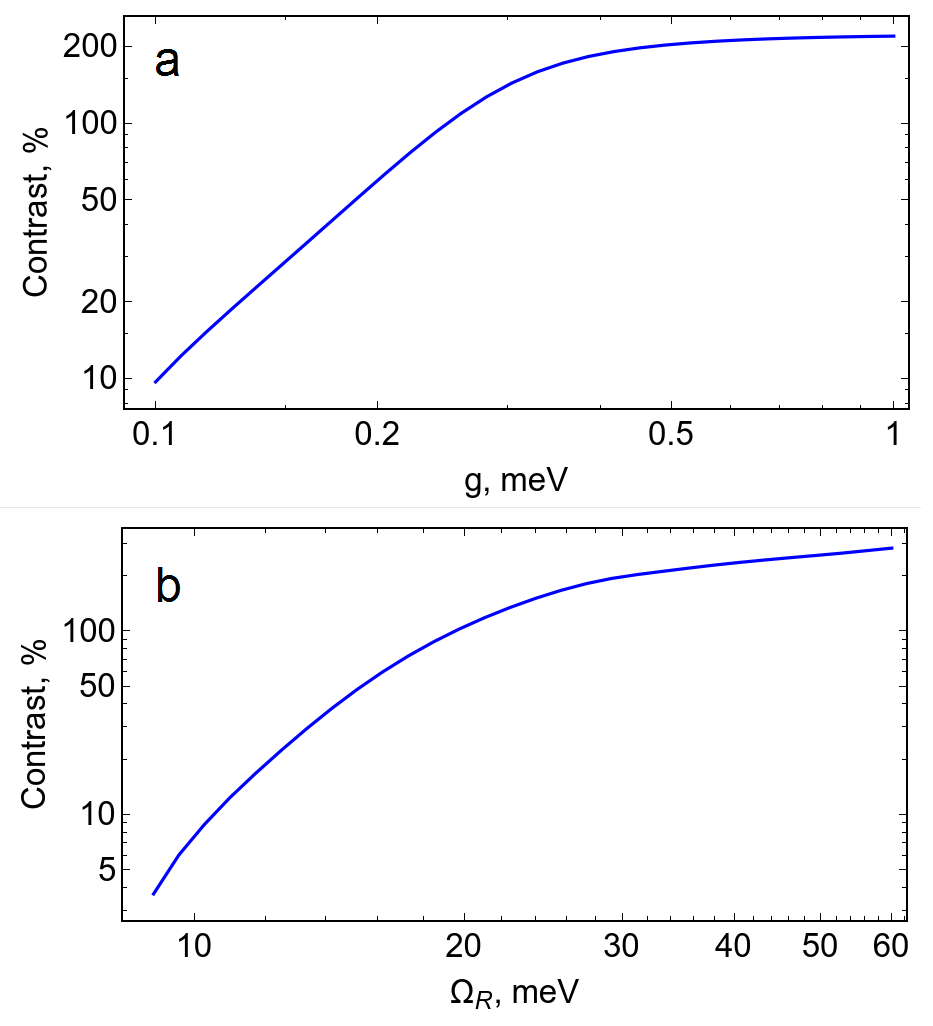}
\caption{ \textbf{a}, The switching contrast as a function of exciton-vibron coupling (${\bf{g}}$) with the fixed exciton-photon coupling $\Omega_{{\rm R}} = 35$~$\rm meV$. \textbf{b}, The switching contrast as a function of exciton-photon coupling ($\bf{\Omega_{{\rm R}}}$) with the fixed exciton-vibron coupling ${g}=0.45$~$\rm meV$. All simulations are done at the fixed pump fluence of $ P=2P_{th}$. Contrast is obtained by integration \textit{E,k} distributions over the ground state momenta range of $\pm 0.2$ $\mu m^{-1}$}
\label{fig:ContrastMB}
\end{figure}

Finally, to peek behind the curtain of the extraordinary nonlinear response of the system, we examine switching contrast defined in the main manuscript as $Contrast = \frac{P_{seed}}{P_{spont}}-1 $ under different exciton-photon ($\Omega_{{\rm R}}$) and exciton-vibron (${\bf{g}}$) coupling constants. Figure \ref{fig:ContrastMB} shows the contrast of switching under \textit{$\langle n \rangle$} = 10 seed photons as a function of coupling constants.

With increasing exciton-photon and exciton-vibron interactions we observe further increase in contrast values followed by saturation regime. The increase of contrast is a quite intuitive result which comes from the form of master equations since both: exciton-photon and exciton-vibron coupling terms in the Hamiltonian Eq.(\ref{ExcitonsCavityHamiltonian}),(\ref{ExcitonsVibronsHamiltonian}) favor stimulated polariton relaxation from higher energy excited states towards the ground state. Further increase of coupling constants does not gain higher switching contrast as the bosonic stimulation is already strong enough to form polariton condensate. The highly nonlinear dynamics of the process saturates polariton population at finite times in this regime of strong coupling terms.

\section{Photon counting of seed pulses}

In this section we provide with calibration data for incident photon states of the seed pulses based on direct method of counting the seed photons transmitted through the sample. The seed beam is focuses onto the sample within wavevector range of $\pm 0.2 \mu m^{-1}$. The light transmitted through the sample is coupled to a single photon avalanche Si photodiode (SAPD) idq100 (ID Quantique) equipped with a 50 $\mu m$ multi-mode fiber. To get rid of the noise we implement time-gated photon counting scheme using time-correlated single photon counting module SPC-160 (Becker \& Hickel GMBH). We employ a standard inverse START-STOP technique to detect the photon events within 1 ns time window allowing for noise-free measurements. A photon detected on SAPD initiates photon counting routine while the signal from the master laser terminates the measurement. We have carried out photon counting experiments at 500Hz repetition rate applying five different seed energies that correspond to 2050, 143, 45, 12 and 3.5 photons per pulse in average, within the linewidth of the ground polariton state. Further on we have quantified integrated photon detection efficiency $(\eta)$, including transmittance of the sample (T), light coupling (K) and quantum efficiency (QE) of the SAPD: $\eta = T\cdot K\cdot QE \approx0.01$. Figure S4 shows an average photon count rate per seed pulse integrated over 60000 pulses (collection time of 120s) versus the average number of photons per pulse incident on the SAPD taking into account integrated detection efficiency $\eta$.

\begin{figure}[h!]
\centering
\includegraphics[width=0.5\textwidth]{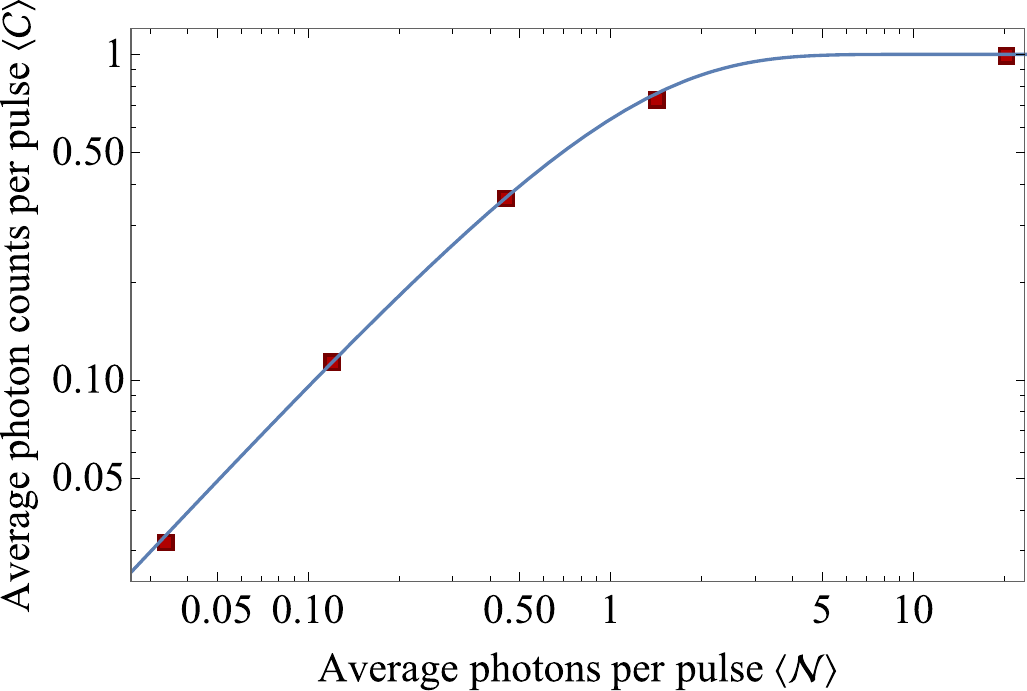}
\caption{ The average photon counts per pulse $\langle C\rangle$ detected by the single photon counting setup within 120s collection time as the function of the average photon numbers per seed pulse $\langle N\rangle$. The blue curve shows model values of $\langle C\rangle$ expected from theory for Poisson-distributed pulses, according analytic expression $\sum_{m=1}^{\infty} \frac{\langle n \rangle^{m}e^{-\langle n \rangle}}{m!} $. }
\label{fig:PhotCounting}
\end{figure}

One can observe a clear linear dependence for the photon counts at the low incident photon numbers $\langle N\rangle \ll1$ that saturates in the vicinity of $\langle N\rangle =1$ with the maximum value of $\langle C\rangle$ equals 1 at $\langle N\rangle \gg1$. The upper limit of $\langle C\rangle=1$ relates to an operation principle of SAPD which generates a constant signal upon photon arrival regardless of the incident photon flux. The overall trend can be quantitatively explained in terms of summation over the probabilities to have at least one photon per seed pulse. Obeying Poisson distribution the sum defines exactly as $\sum_{m=1}^{\infty} \frac{\langle n \rangle^{m}e^{-\langle n \rangle}}{m!} $. We plot the analytic expression in Figure S4 (blue curve). It reproduces our experimental observation in a very precise way that verifies our photon numbers calibration and serves as the independent proof of Poisson distribution of the seed beam.

\clearpage


\end{document}